\documentclass{svmult}
\usepackage{type1cm}
\usepackage{makeidx}
\usepackage{graphicx}
\usepackage{multicol}
\usepackage[bottom]{footmisc}
\usepackage{newtxtext}
\usepackage{newtxmath}
\usepackage{hyperref,url}
\title*{Hybrid-Kinetic Approach: Inertial Electrons}
\author{Neeraj Jain, Patricio A. Mu\~noz and J\"org B\"uchner}
\titlerunning{Hybrid-Kinetic Aproach: Inertial Electrons}
\authorrunning{N. Jain, P.A. Mu\~noz and J. B\"uchner}
\institute{Neeraj Jain 
\at Zentrum f\"ur Astronomie und Astrophysik, Technische Universit\"at Berlin 10623 Berlin, Germany \email{neeraj.jain@tu-berlin.de}
\and Patricio A. Mu\~noz 
\at   Max-Planck-Institut f\"ur Sonnensystemforschung, 37077 G\"ottingen, Germany and \\
  Zentrum f\"ur Astronomie und Astrophysik, Technische Universit\"at Berlin 10623 Berlin, Germany
\and J\"org B\"uchner \at   Max-Planck-Institut f\"ur Sonnensystemforschung, 37077 G\"ottingen, Germany and \\
  Zentrum f\"ur Astronomie und Astrophysik, Technische Universit\"at Berlin 10623 Berlin, Germany}
\begin{document}
\maketitle
\abstract{Hybrid-kinetic simulations describe ion-scale kinetic phenomena in
space plasmas by considering ions kinetically, i.e. as particles, while 
electrons are modelled as a fluid. Most of the existing hybrid-kinetic codes
neglect the electron mass (see chapter 3) for a simplified calculation
of the electromagnetic fields.
There are, however, situations in which delay in the electrons response due to
the electron inertia matters.
This chapter concentrates on hybrid-kinetic simulation models which take the
finite mass of the electron fluid into account.
First a review is given of the history of including the finite electron
mass in hybrid-kinetic models.
Then the equations are discussed which additionally have to
be solved compared to the mass-less hybrid-kinetic models.
For definiteness their numerical implementation without additional approximations is illustrated
by describing a hybrid-kinetic code, CHIEF.
The importance of the consideration of the finite electron mass
are discussed for typical applications (magnetic reconnection, plasma turbulence, collisionless shocks and global magnetospheric
simulations). In particular 
the problem of guide field magnetic reconnection is addressed in some detail.
Possible next steps towards further improvements of hybrid-kinetic
simulations with finite electron mass are suggested.
}

\section{Introduction}
\label{sec:introduction}
The relative scarcity of binary particle collisions in space and astrophysical plasmas makes many of their
plasma processes (like magnetic reconnection, shock waves and turbulence) multi-scale in nature, spanning from the largest,  global fluid- to electron kinetic scales.
In order to investigate those usually non-linear and non-local processes by
computer simulations, one needs to resolve, in principle, the smallest scales necessary for the
processes  and at the same time the large scales of the
system, both in space and time.
While fully kinetic plasma models well describe physical processes down to ion
and electron kinetic scales, they require  prohibitively expensive computational resources to describe at the same time larger scale processes even for
future computational facilities.

The physicists way out of this situation is to understand the processes in parts by
modelling and simulating certain aspects of physics while neglecting
others. This implies the use of a variety of simulation models. A
comprehensive understanding is then obtained by combining various aspects of
these simulations.

In the first part of the book, the basic plasma approximations and models
for the simulation of space and astrophysical plasmas were discussed:
Magnetohydrodynamic (MHD, Chapter 1), Hall-MHD (Chapter 2),
hybrid-kinetic with massless electrons (Chapter 3), gyro-kinetic (Chapter 4), Eulerian Vlasov (Chapter 5)
and Particle in Cell (PIC, Chapter 6) simulation models.
These basic simulation models and codes separately cover all different ranges of
scales and physics.
Kinetic scale processes taking place at ion and electron spatio-temporal scales 
are of critical importance for the mostly collisionless space and astrophysical
plasmas.
For example, collisionless magnetic reconnection, considered to be responsible for
explosive energy release in many space and astrophysical environments including
large scale substorms in the Earth's magnetosphere as well as solar and stellar flares,
is enabled by essentially kinetic scale processes~\cite{Treumann2013}.
The solar wind turbulence (magnetic field fluctuations) is dissipated by kinetic
scales processes~\cite{Bruno2013} and collisionless shock waves essentially
require kinetic scale dissipation processes~\cite{Sagdeev1966}.
Simulations of kinetic scale processes are, therefore,  indispensable to understand
the physics of space and astrophysical plasmas.

Eulerian Vlasov- and semi-Lagrangian PIC-codes (Chapters 5 and 6), in
principle, fully describe ion- and electron scale kinetic plasma processes.
The computational feasibility, however, restricts their use to artificial
physical parameters which reduce, e.g., the gap between electron and ion
scales~	\cite{bret2010}.
Fully kinetic simulations are, therefore, usually carried out for ion-to-electron
mass ratios in the range 25 - 400 while for a full scale separation between
ion- and electron kinetic processes, mass ratios of the order of 1800 would be
necessary.

Furthermore, PIC-code simulations (Chapter 6) suffer from their numerical
shot noise due to a finite number of computationally traced particles
compared to the much larger number of particles in real space and
astrophysical plasma systems~\cite{langdon1970,langdon1979}.
Vlasov-code simulations (Chapter 5) are free from such particle-associated
noise, but they, on the other hand, are computationally equally challenging
as PIC-code simulations due to the necessity to resolve particle distribution
functions on a fine grid in the velocity
space~\cite{BuchnerISSS7:2005, Palmroth2018}.

Gyro-kinetic plasma models (Chapter 4) reduce the dimensions of the velocity space
by one, making simulations this way computationally less demanding.
But they are restricted, e.g., by not resolving the details of the electron and ion
temporal processes~\cite{told2016} and being valid in strongly magnetized
plasmas like in magnetically confining fusion devices and, maybe, in pulsars and
accretion disks.

Hybrid-kinetic plasma simulation models completely neglect kinetic effects associated
with the electrons. This further reduces the computational burden
(see,  e.g.,~\cite{Lipatov2002}).
Such models are, therefore, suitable to study processes for which ion kinetic
effects are important while electron kinetic effects are not.

The majority of existing hybrid-kinetic plasma simulation models simplify their
numerical implementation by assuming a massless (inertia-less) electron fluid.
This assumption allows a direct calculation of the electric field from the electron's
momentum equation without the need of solving  partial differential equations for the electric field (for details, see Chapter 3).
It, however, limits the validity of the model to processes at scales
exceeding by far the electron scales.

Hybrid-kinetic codes with massless electrons have been used 
to simulate global phenomena like the formation of foreshocks and
cavitons by the interaction of the solar wind with the Earth's
magnetosphere~\cite{blanco-cano2009,omidi2013}, day side magnetic
reconnection through the Earth's magnetopause~\cite{hoilijoki2019,pfau-kempf2020},
the formation of filamentary structures in the Earth's
magnetosheath~\cite{omidi2014} and astrophysical
explosions~\cite{winske2007} as well as for specific physics studies of
magnetic reconnection~\cite{Kuznetsova1998,hesse1998,le2016},
plasma turbulence~\cite{franci2016,jain2021} and shock
waves~\cite{weidl2016}).
They contributed to the understanding of space and astrophysical plasma
phenomena at scales of the order of the ion scale or larger.

Effects of the finite electron mass originate from electron scales.
They nevertheless may play an important role for larger scale
plasma phenomena as well.
Laboratory experiments of reconnection have revealed that the
thickness of current sheets maybe as thin as several electron inertial
lengths~\cite{stechow2016}.
At the same time the current sheet thickness can be larger than the
electron gyro-radius.
This allows a fluid description of the electrons. And hybrid-kinetic simulations
with a finite-mass electron fluid have shown that the electron inertia
can dominate when the current sheets  thin down to electron inertial
scale lengths~\cite{hesse1998,Kuznetsova1998}.
In hybrid-kinetic simulations of collisionless plasma turbulence, carried out for
mass-less electrons,  current sheets thin down to the grid scale
\cite{azizabadi2021,jain2021}.  This is not physical. Hence hybrid-kinetic
simulation studies of thinning current sheets must incorporate
electron inertial effects.  Hybrid-kinetic simulations with inertial electrons indeed show the formation of
electron-scale thin current sheets in collisionless plasma turbulence.
Shock steepening to the grid scale has also been observed in hybrid-kinetic simulations (with massless electrons) of
non-stationary shocks ~\cite{hellinger2002}.
Inertial electron hybrid-kinetic simulations showed that the finite mass of the
electrons determines the macroscopic shock
reformation~\cite{yuan2007}.
In particular, it determines the strong phase-space mixing between the upstream
(incoming) and the reflected ions via wave-particle interactions, the
ion thermalization and the growth of plasma waves in the foot and ramp
regions of the shock.

Thus the consideration of the finite electron mass, although it is small,
is important to understand a number of critical phenomena in space and
astrophysical plasmas.

This chapter describes hybrid-kinetic plasma models which take into account
the finite mass of the electrons.
First, in Sect.~\ref{sec:review} a review of the history of the development of
finite-mass-electron codes is given.
In Sect.~\ref{sec:model}) we present the equations which have to be solved.
As an example we discuss in Sect.~\ref{sec:implementation} their numerical
solution by the code CHIEF~\cite{Munoz2018}, their numerical implementation,
the parallelization of the code and its performance and scaling.
In Sect.~\ref{sec:applications_chp9}, importance of the consideration of
the finite electron mass in hybrid simulations are discussed, comparing
results of hybrid-code reconnection simulations with stationary and non-stationary ions.
We also discuss other applications like collisionless shocks, plasma turbulence and global magnetospheric simulations.
Future improvements of related algorithms are discussed in
Sect. \ref{sec:improvements}.

\section{Historical development of finite-electron-mass hybrid-kinetic simulation models}
\label{sec:review}

A number of hybrid-kinetic models have been developed which include different
terms related to the electron inertia under varying degrees of approximations.

First hybrid-PIC models with ions described as macro-particles (PIC
method) and electrons modelled as a fluid were developed together
with the development of fully kinetic-PIC codes~\cite{Forslund1971,Hewett1978}.
Those first hybrid-PIC codes used the Darwin approximation and
solved the electromagnetic field equations separately for the curl-free
and the divergence-free parts of the potentials as well as for the
current density.
In this approximation the electron inertia had to be considered only for calculating
the transverse electron current density.
Practically, hybrid codes based on the Darwin approximation have 
rarely been used, except for 1D cases in which they calculated only
electrostatic potentials.

Then hybrid code algorithms with electron inertia were developed to globally
simulate the entire Earth's magnetosphere~\cite{Swift1996,Swift2001}.
In this approach, the electric field is derived from the electron momentum
equation, the electron velocity is obtained solving Amp\`ere's law and
the magnetic field by solving Faraday's law.
The electron inertia itself was considered only as a correction by taking
into account the electron polarization drift in the (implicit) equation for
the magnetic field, while it was neglected in all other equations.

Later hybrid codes with electron inertia solved for the generalized
electromagnetic fields which satisfy Faraday's-law like equations
(see Sec.~5.7 in~\cite{Lipatov2002},
see also~\cite{Shay1998,Kuznetsova1998}).
These models obtained solutions of the generalized electric and
magnetic field equations by means of predictor-corrector
schemes, which require a staggered grid, or by a trapezoidal
leapfrog algorithm.
The electromagnetic fields were then obtained from
the generalized fields under different approximations.
The magnetic field is calculated solving an elliptic partial
differential equation (PDEs) obtained from the expression
for the generalized magnetic field neglecting electron scale
density variations.
The electric field was then calculated from the generalized
Ohm's law by neglecting the electron inertial term with
time derivatives of the electron fluid velocity~\cite{Kuznetsova1998}. Other authors neglected even the convective electron acceleration
term~\cite{Shay1998}.

These hybrid codes which partially included electron inertial effects
have mainly been used to study collisionless magnetic
reconnection~\cite{Shay1999,Kuznetsova2000,Kuznetsova2001}.
In particular Shay et al. (1998) used an evolution equation for a scalar
electron pressure~\cite{Shay1998} while Kuznetsova et al. (1998) included the 
full electron pressure tensor to take into account the non-gyrotropic 
effects~\cite{Kuznetsova1998} . The latter were shown to play
an important role in possibly balancing the
reconnection electric field.

Another method of calculating the electric field is the solution
of an elliptic PDE for the electric field, obtained by  combining the
Ohm's law with Maxwell's equations (see Sec.~5.2.4 of~\cite{Lipatov2002}).
This approach was utilized for a one-dimensional finite-electron-mass
hybrid code~\cite{Amano2014}.
In that code electric and magnetic fields were obtained by solving the elliptic PDE
and Faraday's law, respectively, and not by calculating for the generalized
electromagnetic fields as in Refs.~\cite{Shay1998,Kuznetsova1998,Lipatov2002}.
Electron inertia effects were considered in the elliptic equation for the electric field
while, still, the electron inertia term was ignored that contains the divergence
of the electric field. This code  used a variable mass ratio  in order to better
model low-density plasma regions~\cite{Amano2014}.
The mass ratio variation was implemented in the code by locally and
temporally varying the electron mass, adjusting it to always satisfy the CFL
condition based on the electron Alfv\'en speed for a given time step (and
grid cell size). This works well as long as the scales of interest are not
too close to the electron inertial scales.

More recently a PIC-hybrid model was developed which implemented
electron inertial terms without any of the approximations used in the
codes before (see above)~ \cite{Munoz2018}.
In Sec.~\ref{sec:implementation} the detailed algorithm and
implementation of this code CHIEF (Code Hybrid with Inertial Electron Fluid)
will be described. Some applications which illustrate the need of
considering the full electron inertia for describing guide field
magnetic reconnection will be discussed
in Sect.~\ref{sec:reconnection}.

Vlasov-hybrid codes directly obtain the ion velocity space distributions
by an Eulerian solution of the Vlasov equations while electrons are
described as an inertial fluid~(see, e.g., \cite{Valentini2007}).
In this approach, the ion Vlasov equations are solved
together with the electron fluid and Maxwell's equations for the
electromagnetic fields.
For the code described by Valentini et al. (2007), a
Helmholtz equation for the electric field was obtained by taking the
curl of Faraday's law, using a generalized Ohm's law and ignoring
density perturbations contained in one of the electron inertial terms~\cite{Valentini2007}.
This code has been applied, e.g., to describe the solar wind
turbulence~\cite{Valentini2008,Valentini2010,Valentini2011} as
well as to other 2D3V and 3D3V plasma
problems~\cite{groselj2017,califano2020,cerri2017}.

Cheng et al. (2013) developed a three-dimensional hybrid-kinetic
code with electron inertia utilizing for the ions a~$\delta f$ method,
i.e. evolving only the variations of the ion distribution function which have to be assumed to be small compared to a given background ~\cite{Cheng2013}.
In this $\delta f$ code the electron inertia was
included in the generalized Ohm's law for the electric field,  calculated
using the Amp\`ere's law and the ion fluid momentum equation.
Such approach was first proposed by Jones et al. in 2003 for
solving the 1D gyrofluid electron equations~\cite{Jones2003}.
It was applied to study the propagation of dispersive Alfv\'en waves
in the coupled Earth's magnetosphere-ionosphere system and electron
acceleration~\cite{Su2004,Su2007} as well as to describe the
Alfv\'en wave dynamics in an Io-Jupiter flux tube~\cite{Su2006}.

\section{Equations to be solved\label{sec:simulation_model}}
\label{sec:model}
Here we describe hybrid-PIC models in which ions are treated as particles and electrons
as a fluid. 
The equations of motion of every ion macro-particle, 

\begin{eqnarray}
        \frac{d\vec{x}_p}{dt}&=&\vec{v}_p\label{eq:xp}          \,\,\,\,\, \text{and} \\
        m_i\frac{d\vec{v}_p}{dt}&=&e(\vec{E}+\vec{v}_p\times \vec{B}), \label{eq:vp}
\end{eqnarray}

where $\vec{x}_p$ and $\vec{v}_p$ are the positions and velocities of ion
macro-particles, respectively, and are solved by a PIC algorithm (see Chapter 6).

The full finite-mass-electron fluid momentum equation is given by

\begin{equation}
        m_e\left[\frac{\partial \vec{u}_e}{\partial t}+(\vec{u}_e\cdot\nabla)\vec{u}_e\right]
        =
        	-e(\vec{E}+\vec{u}_e\times \vec{B})-\frac{1}{n_e} \nabla. \underbar{P}_{e}\,\,, \label{eq:emom}
\end{equation}

where $\vec{u}_e$ and $\underbar{P}_{e}$ are the electron fluid velocity and
the electron pressure tensor, respectively.
The electron density is determined by the quasi-neutrality condition $n_e=n_i=n$.
The electric and magnetic fields, $\vec{E}$ and $\vec{B}$, are related to the plasma current density via Maxwell's equations:
\begin{eqnarray}
	\nabla\times \vec{E}&=&-\frac{\partial \vec{B}}{\partial t}
	\label{eq:faraday}\\
	\nabla\times \vec{B}&=&\mu_0ne(\vec{u}_i-\vec{u}_e)
	\label{eq:ampere}
\end{eqnarray}
Here $\vec{u}_i$ is the ion fluid velocity obtained by the first order velocity moment of the ion distribution
function.
From Eqs. \ref{eq:emom} and \ref{eq:faraday}, we obtain an evolution equation for the
generalized vorticity $\vec{W}=\nabla\times\vec{u}_e-e\vec{B}/m_e$.
\begin{eqnarray}
\frac{\partial\vec{W}}{\partial t}
  &= &\nabla\times (\vec{u}_e\times \vec{W})-\nabla\times\left(\frac{\nabla p_e}{m_en_e}\right)\label{eq:curl_emom}\\
  p_e&=&n_eT_e\label{eq:eos}
\end{eqnarray}
For simplicity, usually the electron pressure is assumed to be a scalar quantity $p_e$.
For a uniform temperature $T_e$ the last term in Eq.~\ref{eq:curl_emom} vanishes.

Substituting $\vec{u}_e$ in the expression of $\vec{W}$ from Ampere's law
one obtains an elliptic equation for the magnetic field,
\begin{eqnarray}
	\frac{1}{\mu_0e}\nabla\times\left(\frac{\nabla\times\vec{B}}{n}\right)+\frac{e\vec{B}}{m_e}&=&\nabla\times\vec{u}_i-\vec{W}\label{eq:elliptic_b}
\end{eqnarray}

Eqs.~\ref{eq:xp}--\ref{eq:elliptic_b} represent a complete hybrid-PIC plasma model
without any approximation on the electron inertial terms in the electron momentum
equation.
Note that in conventional hybrid codes, which neglect the electron inertia,
the terms on the left hand side (LHS) of Eq. \ref{eq:emom} are neglected
allowing a direct calculation of $\vec{E}$ from this equation.
As a result the $\nabla \times \vec{u}_e$ term  in the expression of $\vec{W}$
disappears and Eq. (\ref{eq:curl_emom}) becomes an evolution equation
for the magnetic field, i.e. Eq. (\ref{eq:elliptic_b}) does not have to be solved at all.

In conventional hybrid codes with electron inertia, the electric and magnetic fields
are obtained from Eqs.~\ref{eq:emom} and \ref{eq:curl_emom} by usually
neglecting some of the terms.
Substituting for  $\vec{u}_e$ from Eq.~\ref{eq:ampere} and neglecting terms proportional to $\partial \vec{u}_i/\partial t$ and $\partial n/\partial t$, the LHS of Eq.~\ref{eq:curl_emom} can be written entirely in terms of the time derivative of the magnetic field. The neglect of  $\partial \vec{u}_i/\partial t$ and $\partial n/\partial t$ in Eq.~\ref{eq:curl_emom} is justified for a large mass ratio $m_i/m_e$. However these approximations are not necessarily valid in all situations of interest and need to be separately justified, especially for artificially small mass ratios $m_i/m_e$.
Neglecting electron scale variation of the density, the LHS of that equation can further be simplified
to $\partial/\partial t[e(\vec{B}-d_e^2\nabla^2\vec{B})/m_e]$, where $d_e=c/\omega_{pe}$ is the electron inertial length. The electric field is calculated directly from Eq.~\ref{eq:emom} by neglecting $\partial\vec{u}_e/\partial t$, which is clearly inconsistent with  keeping this term to obtain Eq.~\ref{eq:curl_emom}.

In another approach, the electric field is calculated from Eq.~\ref{eq:emom}
by explicitly evaluating the time derivative term $\partial\vec{u}_e/\partial t$~\cite{Munoz2018}.
For this purpose Munoz et al.  in 2018 used the heaviness of the ions to assume that
the ions quantities do not change much during a single time step  i.e. during a
fraction of the electron gyro-period~\cite{Munoz2018}.
Thus for an update of ion positions and velocities by a single time step $dt$,  $\vec{u}_e$ at $t=t_0+dt$ and $t=t_0+2\,dt$
can be obtained by advancing Eq.~\ref{eq:curl_emom}, first from $t=t_0$ to $t=t_0+dt$
and then from $t=t_0+dt$ to $t=t_0+2\,dt$.
This allows the calculation of  $\partial\vec{u}_e/\partial t$ by a central difference scheme.
In the second of the two time steps, the ion current and density from the previous time step
can be used.
Alternatively the electric field can be calculated by solving an elliptical PDE which one
obtains by taking the curl of the Faraday's law (Eq.~\ref{eq:faraday}), using
Amp\`ere's law (Eq.~\ref{eq:ampere}) and substituting for
$\partial\vec{J}/\partial t$ from the generalized Ohm's law.
The numerical solution of the resulting elliptical PDE in more than one dimension is,
however, much more involved than that of the elliptic PDE for the magnetic field
(Eq.~\ref{eq:elliptic_b}).
The difficulty arises from the fact that one cannot use a Poisson equation to
substitute for $\nabla\cdot\vec{E}$, which yields cross-derivative terms
in the equation.
In case of Eq.~\ref{eq:elliptic_b}, the condition $\nabla\cdot\vec{B}=0$
simplifies the calculations.


\section{Numerical Implementation}
\label{sec:implementation}

For definiteness we illustrate, how a hybrid-PIC simulation model
with a finite-mass electron fluid without additional approximations
(see Sec.~\ref{sec:model}) works, by describing the model's numerical implementation
in a code called CHIEF (Code Hybrid with Inertial Electron Fluid).
CHIEF combines elements of the PIC code ACRONYM
to describe the ions as particles
(see, e.g.~\cite{Kilian2012} \footnote{\url{http://plasma.nerd2nerd.org/}})
with an EMHD code which solves the electromagnetic equations coupled
with an inertial electron fluid~(see, e.g., \cite{Jain2006}).
Both codes have separately been tested and independently used to
simulate different aspects of magnetic reconnection~\cite{Jain2014b,Munoz2016a},
plasma instabilities~\cite{Jain2014e,Munoz2014a},
particle acceleration~\cite{Burkart2010a}, shocks~\cite{Kilian2015},
wave coupling~\cite{Ganse2014},
resonant wave-particle interactions~\cite{Schreiner2017} etc.
These simulations have helped to  clean up both the codes off numerical
errors.

\subsection{Ions as particles}
\label{sec:ions}

In the ion-related (PIC-) part of the code the macro-particle positions and
velocities at the $N$th time step are advanced for the electric and magnetic
field values known at the time step $N$ and defined on a staggered grid,  the Yee-lattice grid~\cite{Yee1966}.
For this the fields are first interpolated from the grid to
the (macro-) ions positions with a weighting given by a
shape function $S(\vec{x}-\vec{x}^{\;p})$:

\begin{eqnarray}\label{eq:interpolation_em}
    \vec{E}^{\;p}& = &\vec{E}(\vec{x}^{\;p})=\int \vec{E}(\vec{x})\,S(\vec{x}-\vec{x}^{\;p})\,d\vec{x},                              \\
    \vec{B}^{\;p} &= &\vec{B}(\vec{x}^{\;p}) = \int \vec{B}(\vec{x})\,S(\vec{x}-\vec{x}^{\;p})\,d\vec{x}, \label{eq:interpolation_em2}
  \end{eqnarray}

where the super-script $p$ indicates the location of each (macro-)particle.
Then, the ions are moved by a second-order accurate leap-frog algorithm.
This means that at each time step $N$ the ion velocities are advanced from
the  half time step $N-1/2$  to the half time step $N+1/2$ and the ion
positions from time step $N$ to time step $N+1$ by using the
discretized version of Eqs.~\ref{eq:xp}
and~\ref{eq:vp} (see, e.g., Sec.~4.3 of~\cite{Birdsall1991}):

  \begin{eqnarray}
    \label{eq:xp_num}  \vec{x}^{p,N+1}  & = &\vec{x}^{p,N} + \vec{v}^{p,N+1/2}\Delta t ,                                                                                             \\
    \label{eq:vp_num} \vec{v}^{p,N+1/2} & = &\vec{v}^{p,N-1/2} + \frac{e\Delta t}{m_i}\left( \vec{E}^{p,N}+\frac{\vec{v}^{p,N+1/2}+\vec{v}^{p,N-1/2}}{2}\times \vec{B}^{p,N}\right)
  \end{eqnarray}

 Note that the first equation is explicit for $\vec{x}^{p,N+1}$ while the
 second is implicit  for $\vec{v}^{p,N+1/2} $. The ion velocity is advanced 
 via a Boris method~\cite{Boris1970}, i.e. a rotation of $\vec{v}^{p,N+1/2}$
 (i.e., the magnitude of the vector stays constant), making Eq.~\ref{eq:vp_num}
 also explicit.

Because of the staggered position- and velocity updates the sources of the
electromagnetic field, the ion number density $n_i$
and ion current density $\vec{\jmath}_i$, are computed at
time steps indexed $N+1$ and $N+1/2$, respectively.
The deposition of both the ion number and current density onto the grid is done via an interpolation scheme using the same shape function  as for the electromagnetic field interpolation from the grid to the macro-particles position (Eqs.~\ref{eq:interpolation_em}-\ref{eq:interpolation_em2}):
  \begin{eqnarray}\label{eq:interpolation_n_j}
                n_i^{N+1}(\vec{x}) & = &\sum_p N_p\,S(\vec{x}-\vec{x}^{\;p,N+1}) ,                                               \\
                \vec{\jmath}_i^{\;N+1/2}(\vec{x}) & =& \sum_p N_p\,\vec{v}^{\;p,N+1/2} S(\vec{x}-\vec{x}^{\;p,N+1}).\label{eq:interpolation_n_j2}
  \end{eqnarray}
The use of the same shape function in Eqs.~\ref{eq:interpolation_em}  and \ref{eq:interpolation_n_j}
avoids self-forces and preserves the global momentum (see, e.g., Secs.~8.5-6 
in~\cite{Birdsall1991} and Sec 5.3.3 of~\cite{Hockney1988} and
~\cite{Brackbill2016} for fully-kinetic PIC codes or Sec.~4.5.2 in
~\cite{Lipatov2002} for hybrid-PIC codes).

\subsection{Electron fluid}
\label{sec:electrons}

The electron density $n_e^{N+1/2}$ is set equal to the ion density $n_i^{N+1/2}=(n_i^{N}+n_i^{N+1})/2$,
i.e. $n_e^{N+1/2}$ is obtained by applying the quasi-neutrality condition. If a
constant (not evolving in time) electron temperature  $T_e$ is assumed the electron
pressure $p_e^{N+1/2}=n_e^{N+1/2}k_BT_e$ can be calculated via the equation of
state, Eq.~\ref{eq:eos}.


\subsection{Electromagnetic fields}
\label{sec:emfields}

Once the sources of the electromagnetic fields (Eqs.~\ref{eq:interpolation_n_j}-\ref{eq:interpolation_n_j2})
are known on the Yee lattice, the electromagnetic fields are updated by using the electron fluid
equations  and Maxwell's equations.
The electric and magnetic fields can be updated, e.g., by advancing the generalized vorticity
$\vec{W}$ in Eq.~\ref{eq:curl_emom} from time step $N$ to  $N+1$
using a flux-corrected transport algorithm like the one provided by the LCPFCT 
package~\footnote{\texttt{\url{http://www.nrl.navy.mil/lcp/LCPFCT}}}.
The LCPFCT package   solves the  
generalized continuity equations and has been developed by the U.S. Naval Research 
Laboratory.
Note that the use of the  LCPFCT algorithm is advantageous to resolve steep gradients.

The full update to the next time step ($\vec{W}^{N+1}$) requires  $\vec{W}^{N+1/2}$ and $\vec{u}_e^{N+1/2}$, which are not known yet.
Because of this, the field solver has to first advance the  generalized
vorticity $\vec{W}^N$ by half a time step (with the input quantities  $n_e^{N+1/2}$, $p_e^{N+1/2}$, $\vec{u}_e^{N}$,  $\vec{W}^N$, $\vec{B}^N$) in order to estimate $\vec{u}_e^{N+1/2}$,  $\vec{W}^{N+1/2}$, and $\vec{B}^{N+1/2}$,  by solving the discretized version of Eq.~\ref{eq:curl_emom}:

\begin{eqnarray}\label{eq:curl_emom_discrete}
        \left.\frac{\partial \vec{W}}{\partial t}\right|_{N\to N+1/2}
                & = \vec{\nabla}\times\left [\vec{u}_e^N\times \vec{W}^N\right]-\vec{\nabla}\times\left(\frac{\vec{\nabla} p_e^{N+1/2}}{m_en_e^{N+1/2}}\right).
\end{eqnarray}


A three-dimensional flux-corrected transport algorithm for solving continuity
equations (LCPFCT) is used to solve Eq. ~\ref{eq:curl_emom_discrete}.

The solution for $\vec{B}^{N+1/2}$ of the discretized version of the 
elliptic Eq.~\ref{eq:elliptic_b} can be obtained by solving

\begin{equation}
	\vec{\nabla}\times\left(\frac{\vec{\nabla}\times\vec{B}^{N+1/2}}{\mu_0en^{N+1/2}}\right)+\frac{e\vec{B}^{N+1/2}}{m_e} =\vec{\nabla}\times\vec{u}_i^{N+1/2}-\vec{W}^{N+1/2}. \label{eq:b_half}
\end{equation}

The solution of Eq.~\ref{eq:b_half} allows to obtain $\vec{B}^{N+1/2}$ from $\vec{W}^{N+1/2}$.
The discretization  of this elliptical PDE reveals a system of algebraic equations
which can be solved, e.g., by a general purpose scalable linear multi-grid solver contained
in the multigrid-library
HYPRE~\cite{Falgout2002,falgout2006}~\footnote
{\texttt{\url{https://hypre.readthedocs.io/en/latest/ch-intro.html}}}.


Next the components of $\vec{u}_e^{N+1/2}$ can be obtained by solving the
discretized Amp\`ere's law, Eq.~\ref{eq:ampere}.

\begin{eqnarray}
	\vec{u}_e^{N+1/2}=\vec{u}_i^{N+1/2}-\frac{\vec{\nabla}\times\vec{B}^{N+1/2}}{\mu_0en_e^{N+1/2}}\label{eq:ue_half}.
\end{eqnarray}

At this point all relevant quantities are known at time step
$N+1/2$: $\vec{u}_e^{N+1/2}$, $\vec{B}^{N+1/2}$ and $\vec{W}^{N+1/2}$.
These quantities can now be advanced to time step $N+1$ by solving the
following equations.

\begin{eqnarray}
        \left.\frac{\partial \vec{W}}{\partial t}\right|_{N \to N+1}
        & = & \vec{\nabla}\times\left [\vec{u}_e^{N+1/2}\times \vec{W}^{N+1/2}\right]\nonumber \\
                        & \qquad -&\vec{\nabla}\times\left(\frac{\vec{\nabla} p_e^{N+1/2}}{m_en_e^{N+1/2}}\right) \label{eq:w_full} \\
        \vec{\nabla}\times\left(\frac{\vec{\nabla}\times\vec{B}^{N+1}}{\mu_0en_e^{N+1/2}}\right)+\frac{e\vec{B}^{N+1}}{m_e} & = &\vec{\nabla}\times\vec{u}_i^{N+1/2}-\vec{W}^{N+1}\label{eq:b_full}                                                                                                \\
        \vec{u}_e^{N+1}                                                                                                     & = &\vec{u}_i^{N+1/2}-\frac{\vec{\nabla}\times\vec{B}^{N+1}}{\mu_0en_e^{N+1/2}}\label{eq:ue_full}
\end{eqnarray}


 Eq.~\ref{eq:w_full}  can be solved by using already mentioned three-dimensional
 LCPFCT package developed by the U.S. Naval Research Laboratory and
 Eq.~\ref{eq:b_full} by utilizing the HYPRE library package.
This way one obtains $\vec{W}^{N+1}$,  $\vec{B}^{N+1}$ and $\vec{u}_e^{N+1}$.
Note that the ion quantities ($\vec{u}_i$ and $n_i$) are used at time steps $(N+1/2)$,
same as for the first half time step.
This is justified by the heavy ions whose mass is much larger compared to that
of the electrons.



For the calculation of the electric field, equations \ref{eq:ampere}, \ref{eq:curl_emom} and \ref{eq:elliptic_b}  are again advanced from 
time step $N+1$ to $N+2$ but using the ion quantities at the time index $N+1/2$  (i.e., assuming that they are
temporally fixed).
This reveals $\vec{B}^{N+2}$, $\vec{W}^{N+2}$ and $\vec{u}_e^{N+2}$.
Now, the electric field can be calculated at time step $N+1$ using the discretized version of the 
generalized Ohm's law, Eq.~\ref{eq:emom}:

\begin{eqnarray}
	\vec{E}^{N+1}&=&-\frac{m_e}{e}
	\left[
	\frac{\vec{u}_e^{N+2}-\vec{u}_e^{N}}{2\, \Delta t}
	+(\vec{u}_e^{N+1}\cdot\vec{\nabla})\vec{u}_e^{N+1}
	\right]\nonumber \\
	&\qquad -&\vec{u}_e^{N+1}\times\vec{B}^{N+1}
	-\frac{\vec{\nabla} p_e^{N+1}}{en_e^{N+1}}. \label{eq:E_at_NP1}
\end{eqnarray}

The value of $p_e^{N+1}=n_e^{N+1}k_BT_e$ are provided by the equation
of state, Eq.~\ref{eq:eos}, in the same way as before. Note that 
$p_e^{N+1}=p_e^{N+1/2}$ and $n_e^{N+1}=n_e^{N+1/2}$ are 
assumed.

Finally, the values of $\vec{E}^{N+1}$ and $\vec{B}^{N+1}$ are available to
move the ions further. This ends a time step and the cycle is repeated.

Note that the model and numerical implementation as presented in
Sects. \ref{sec:simulation_model} and \ref{sec:implementation}
do not contain resistive terms.
Finite resistivity is usually included in hybrid codes with massless electron
fluid to damp whistler waves propagating at the grid scale.
This is necessary since  the whistler phase
velocity for a massless electron fluid increases  with the wave number without bounds.
As a result the time steps have to become very small with increasing
grid resolution to further satisfy the CFL (Courant) condition and still
resolve the whistler wave propagation.

The phase speed of whistlers for an inertial electron fluid, on the other hand,
is limited due to the finite electron mass.
Therefore, the use of finite resistivity is not necessary for keeping the
hybrid codes with massive electrons stable (see Sects. \ref{sec:simulation_model}
and \ref{sec:implementation}). It can, however, be implemented in the code
to model collisional resistivity in cases of interest.
The PIC-typical particle shot noise can be reduced in two different ways.
One is the use of higher (at least second) order shape functions for
the current density deposition from particles to grid and vice-versa
to calculate the electromagnetic fields at the ion position.
A second possibility in the code is to apply a standard binomial filter
on the current  deposition, smoothing out the ion current density to be used 
as a source term in the Maxwell solver.

\subsection{Code parallelization and performance}
\label{sec:parallelization}

\begin{figure}[ht!]
		\includegraphics[width=0.99\textwidth]{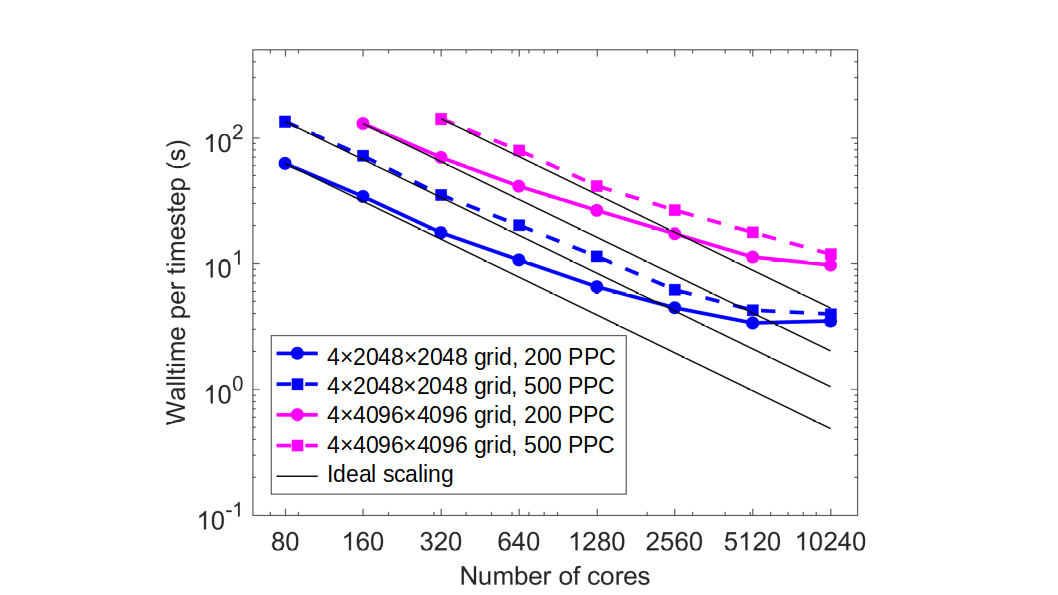}
		\caption{Strong scaling results (run time as a function of compute cores)
		for the CHIEF code using up to 10240 cores (256 nodes with 40 compute cores
		per node). Reprinted from \cite{Jain2022}, with the permission of AIP Publishing. 
		\label{fig:strongscaling}}
\end{figure}

The CHIEF code is parallelized and optimized for a high efficiency
on state-of-the-art supercomputer clusters using a domain decomposition
approach  implemented based on Message Passing Interface (MPI)
commands.
The three-dimensional LCPFCT algorithm~\cite{Zalesak1979} used to
solve  Eqs. \ref{eq:curl_emom_discrete} and \ref{eq:w_full} in a massively
parallel way is computationally much more efficient than the
one-dimensional LCPFCT~\cite{Boris1993} algorithm used in the non-parallel code
(see, e.g.~\cite{Munoz2018}).
The scalable linear solvers of the HYPRE package  using multigrid methods allows
to efficiently solve the elliptic equations~\ref{eq:b_half} and \ref{eq:b_full}.
As a result it was shown that a massively parallelized CHIEF code performs
well for two- as well as for three-dimensional spatial grids.
As an example, Fig.~\ref{fig:strongscaling}  depicts the strong scaling behaviour
of the code obtained for quasi-two dimensional grids on the HPC System COBRA
of the Max-Planck-Computing and Data Facility (MPCDF) in Garching (Germany).
The COBRA cluster used for the scaling shown in~Fig.~\ref{fig:strongscaling}
consists of Intel Xeon Gold 6148  (``Skylake'' architecture) processors with
40 cores per node, operated at the nominal frequency of 2.4 GHz (Turbo mode
disabled). The nodes are connected through a non-blocking 100 Gb/s OmniPath
interconnect with a non-blocking fat-tree topology. Further Intel 19.1.3 C++ and
Fortran compilers were used as well as the Intel MPI 2019.7,
Intel MKL 2020.4, Blitz++ 1.0.1, and the HYPRE 2.22 libraries.
Fig.~\ref{fig:strongscaling} shows the run time per time step (average of the first
three time steps in 4 different runs) as a function of the number of computing
cores (MPI tasks) for two different grid sizes  ($4\times 2048\times 2048$, and
$4\times 4096\times 4096$)  up to 10240 cores (256 nodes) with 200 and 500 particles per cell. The black lines indicate ideal strong scaling, taking the computation time on the minimum number of nodes with 768 GB RAM which could run the calculations as the reference.
For the larger grid size ($4\times 4096\times 4096$) with 500 particles per cell the code
maintains a good strong scalability, up to roughly 2560 cores.
It is able to at least decrease the run time down to ca. 12 s per time step when
using 10240 cores.
As expected, deviation from close-to-ideal scaling occurs at smaller core counts for smaller
grids or smaller number of particles per cell.

The code CHIEF also performs well on the computer clusters of the Max-Planck-Institute
for Solar System Research in G\"ottingen (Germany) and the computer clusters
of the Technical University Berlin.

\section{Applications}
\label{sec:applications_chp9}

In the following we discuss the importance of the consideration of the
finite electron mass in hybrid-kinetic codes for four typical space and
astrophysical plasma processes: collisionless guide-field magnetic
reconnection (Sec. \ref{sec:reconnection}), collisionless plasma turbulence
(Sec. \ref{sec:turbulence}), collisionless shocks (Sec. \ref{sec:shocks}) and
global simulations of collisionless magnetospheric plasmas
(Sec. \ref{sec:global}).
In particular we  discuss at some length the importance of the
electron inertia for guide field magnetic reconnection by
applying the above described code CHIEF.

\subsection{Magnetic reconnection}
\label{sec:reconnection}

The magnetic reconnection phenomenon was historically first
investigated in the framework of MHD model (see Chapter 1).
Later the consequences of the decoupling of the electron and ion fluids
were addressed by Hall-MHD models and simulations
(see Chapter 2).
The role of kinetic ions was  investigated by hybrid-kinetic codes
neglecting the electron mass (see Chapter 3).
Finally, in the end of the 1990s, the finite electron mass was taken into account in hybrid-kinetic codes to describe magnetic reconnection.

Shay et al. (1998), e.g., retained electron inertial terms but only in
the equation of  the generalized magnetic field and not in that of the generalized electric field~\cite{Shay1998,Shay1998b}.
By simulating this way reconnection between two magnetic flux
bundles with a zero net-current and a small overlap in order to
trigger reconnection, these simulations identified a small region
centered around the X-line and extending toward the separatrices of
reconnection as a region in which electron inertial effects dominate.
These regions with a size of the order of an electron inertial length
are much smaller than the Hall region. In these regions, sharp gradients of
the electron velocity develop which are susceptible to electron shear
flow instabilities.

Kuznetsova et al. (1998) retained the convective derivative of
the electron fluid velocity and also included the non-gyrotropic electron pressure tensor terms in the expression of the
generalized electric field~\cite{Kuznetsova1998,Kuznetsova2000,Kuznetsova2001}.
Starting with a Harris current sheet equilibrium without an out-of-plane
directed guide magnetic field and adding a small perturbation to trigger reconnection, 
they  found that the electron inertial term is subdued by the pressure tensor
term in supporting the steady-state reconnection electric field~\cite{Kuznetsova1998}. Finite electron inertia, however, contributes to the reconnection
electric field in a non-steady fashion for steep gradients at
the electron inertial length scale.
It was suggested that the electron inertia term may dynamically dominate
the influence of the non-gyrotropic pressure tensor terms at electron
time scales and for reconnection in finite guide magnetic field.

Almost a decade later first two-dimensional fully-kinetic PIC-code simulations
with open boundary conditions showed that reconnection is, indeed, inherently
non-stationary due to the development of secondary instabilities of the electron layer~\cite{daughton2006}.
Three dimensional fully-kinetic PIC simulations of magnetic reconnection
with guide field confirmed the growth of secondary instabilities of the
electron current layer~\cite{daughton2011}. It was shown that the electron
inertia is necessary for driving these instabilities~\cite{che2011}.

Hybrid-kinetic simulation models with finite electron inertia are able to
describe not only the electron-inertia related instabilities of the thin
electron layers as fully kinetic simulations but also their consequences for
the larger (ion) scale dynamics of reconnection. The algorithm of the
CHIEF code which implements electron inertia without any approximation
(see Sec.~\ref{sec:implementation}) is appropriate for such studies including
the description of the electromagnetic fluctuations and turbulence in reconnection
regions.

\subsubsection{Electromagnetic fluctuations in reconnection regions}
\label{subsec:vineta}

{\it In-situ} space observations of magnetic reconnection have disclosed that
electromagnetic and electrostatic fluctuations are generated in and around
reconnecting current sheets (see, e.g., ~\cite{eastwood2009,zhou2009,retino2007}.
Laboratory experiments~\cite{ji2004,fox2010,inomoto2013,dorfman2014,kuwahata2014,stechow2016}
revealed electromagnetic fluctuations in the lower hybrid frequency range~\cite{zhou2009,ji2004,stechow2016} as well as in the ion
acoustic~\cite{inomoto2013}, Trivelpiece-Gould~\cite{fox2010} and Langmuir
wave frequency ranges~\cite{gekelman1984}.
In the~\textsc{VINETA}.II experiments~\cite{stechow2016}, reconnection
was driven in a linear laboratory device (cf. the cartoon in Fig.~\ref{fig:x-point})
by applying an external time varying magnetic field to a plasma immersed in  a uniform and constant guide magnetic field 
$B_{g}$ in the $z$ (out-of-plane) direction and a figure-eight X-point
field $\vec{B}_{\perp}(t)$ in the perpendicular $x$-$y$ plane.
In the resulting reconnecting electron current sheet with a half
thickness $\sim 5\,d_e \approx \rho_i/2 \approx 25 \rho_e$, 
electromagnetic fluctuations in the lower hybrid frequency range were
generated.
At these scales, ions behave as particles while electrons can be treated
as a fluid whose inertia is important.
A hybrid-kinetic plasma model with kinetic ions and an inertial electron fluid
like CHIEF is, therefore, suitable to describe the fluctuations near the
lower hybrid frequency as observed in space as well as in VINETA.II.

\subsubsection{EMHD simulation of guide-field magnetic reconnection with
finite electron mass but with immobile ions}
\label{sec:immobile}

Before presenting hybrid-PIC CHIEF simulation results  of guide-field
magnetic reconnection, let us demonstrate the role of the electron inertia by 
first using the framework of an electron-magnetohydrodynamic (EMHD)-model.
EMHD models are the simplest hybrid models since they allow to study
the effects of the electron inertia while neglecting the influence of ions,
i.e. considering the ions to be immobile.
EMHD simulations of guide-field magnetic reconnection were carried out
by Jain et al.  in 2017~\cite{jain2017} for a configuration similar to that of the
VINETA.II  experiment~\cite{stechow2016}.
Several consequences of the finite electron mass for guide-field reconnection
could be revealed already in the framework of the restricted EMHD description:
the formation of a thin electron scale current sheet, development of electron inertial instabilities
and the generation of electromagnetic fluctuations.

\begin{figure}[!ht]
  \centering{\includegraphics[width=0.7\textwidth]{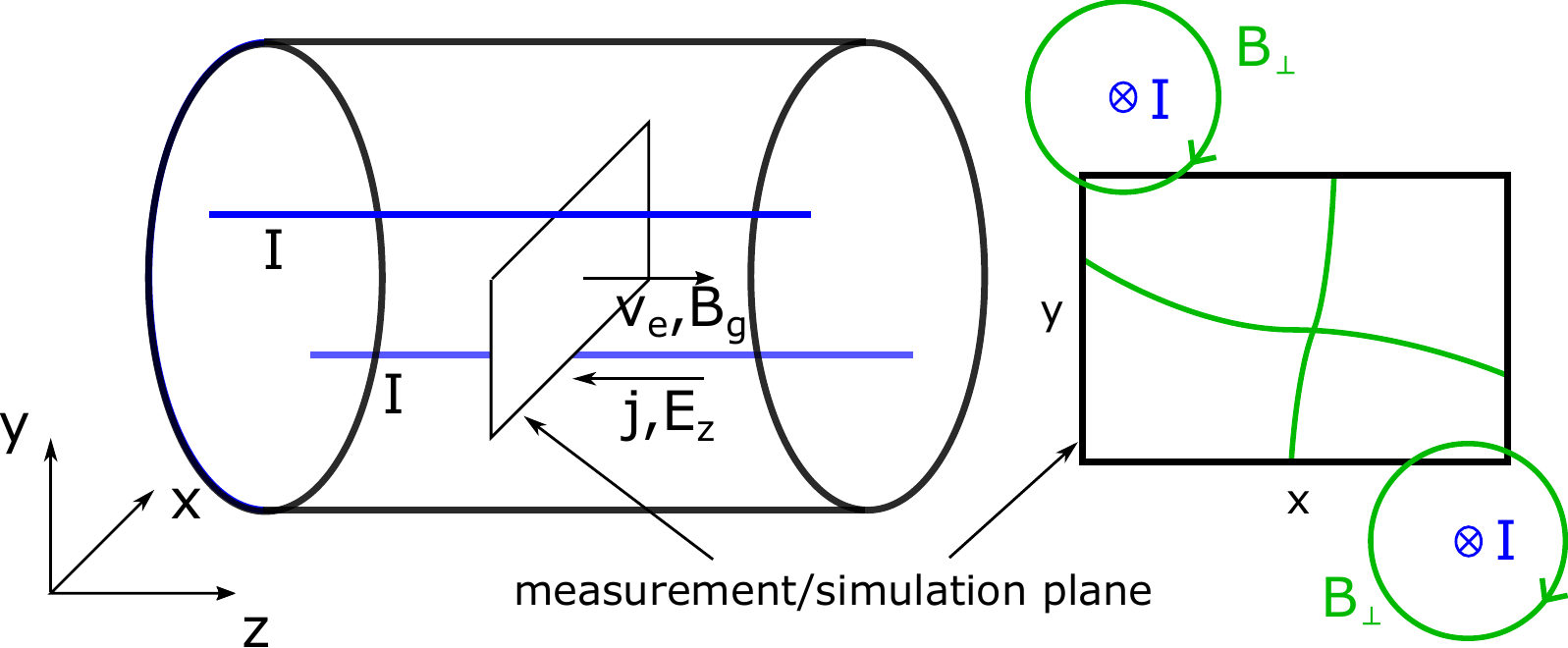}}
  \vspace{0.2in}\\
  \centering{\includegraphics[width=0.7\textwidth]{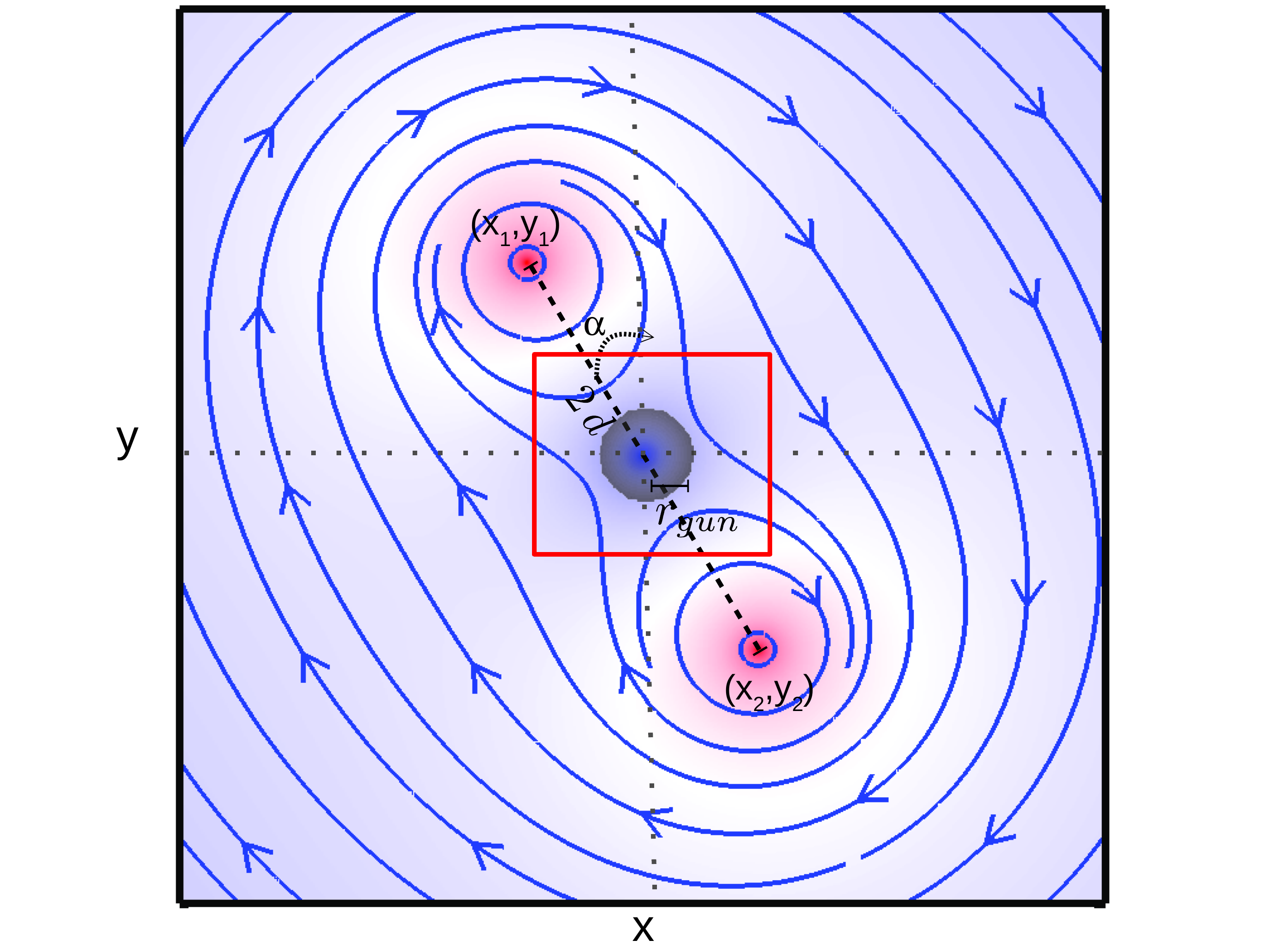}}
  \caption
  {Top: experiment overview and measurement plane. Bottom: simulation plane -  Color-coded magnitude (red: high and blue: low) and field lines (blue)
of the perpendicular external magnetic field $\vec{B}_{\perp}^{ext}$ produced by  two long wires at the labeled positions ($x_1,y_1$) and ($x_2,y_2$)
carrying current in the negative z-direction.
The shaded area at the center of the simulation box (red square) represents the extent of the external electric field $E_z^{ext}$
(Reproduced from~\cite{Jain:2017-Vineta} with the permission of AIP Publishing).}
\label{fig:x-point}
\end{figure}

The EMHD simulations were initialized with a uniform, motionless plasma
embedded in an externally imposed magnetic field $\vec{B}^{ext}=\vec{B}_{\perp}^{ext}+B_g\hat{z}$ with a guide field $B_g=15\,$mT. The perpendicular magnetic field is created by two infinitely long wires separated by a distance $2\,d=30$ cm and carrying a current $I_0$=2 kA along the negative $z$-direction (see Fig. \ref{fig:x-point}). To match the experiment geometry, the wire positions are rotated by $\alpha =30^\circ$ with respect to the $y$ axis.
In the experiments, the plasma current is extracted from a localized electron source (the plasma gun) with a radius $r_{gun}=6$ mm by the large scale electric field $E_z$, which in the simulations is approximated by $\vec{E}_z^{ext}=-E_{z0}\exp\left(-\frac{x^2+y^2}{2r_{gun}^2}\right)\,\hat{z},$
with $E_{z0}=10\,$V/m. The simulation box size,  10 cm $\times$ 10 cm $\times$ 10 cm, corresponds to the experimental measurement area. Note that the parallel wires are outside the simulation domain (see Fig. \ref{fig:x-point}). The grid resolution in each direction is 1.35\,mm.  Simulations are run for $3.6\,\mu$s in steps of 0.2\,ns. The collisional resistivity is taken to be zero. The simulation results are presented in normalized variables: Magnetic fields are normalized by the edge magnetic field produced by the experimental current sheet ($B_{edge}=0.8$ mT), time by the inverse electron cyclotron frequency in this magnetic field ($\omega_{ce}^{-1}=(eB_{edge}/m_e)^{-1}$=7.2 ns), distances by the electron inertial length ($d_e$=2.7 mm), and the current density by $n_0ev_{Ae}=B_{edge}/\mu_0 d_e$=235.8 kA/m$^2$.

\begin{center}
        \begin{figure}[!ht]
                \includegraphics[clip,width=0.49\textwidth,trim=1cm 0.3cm 1cm 0.2cm]{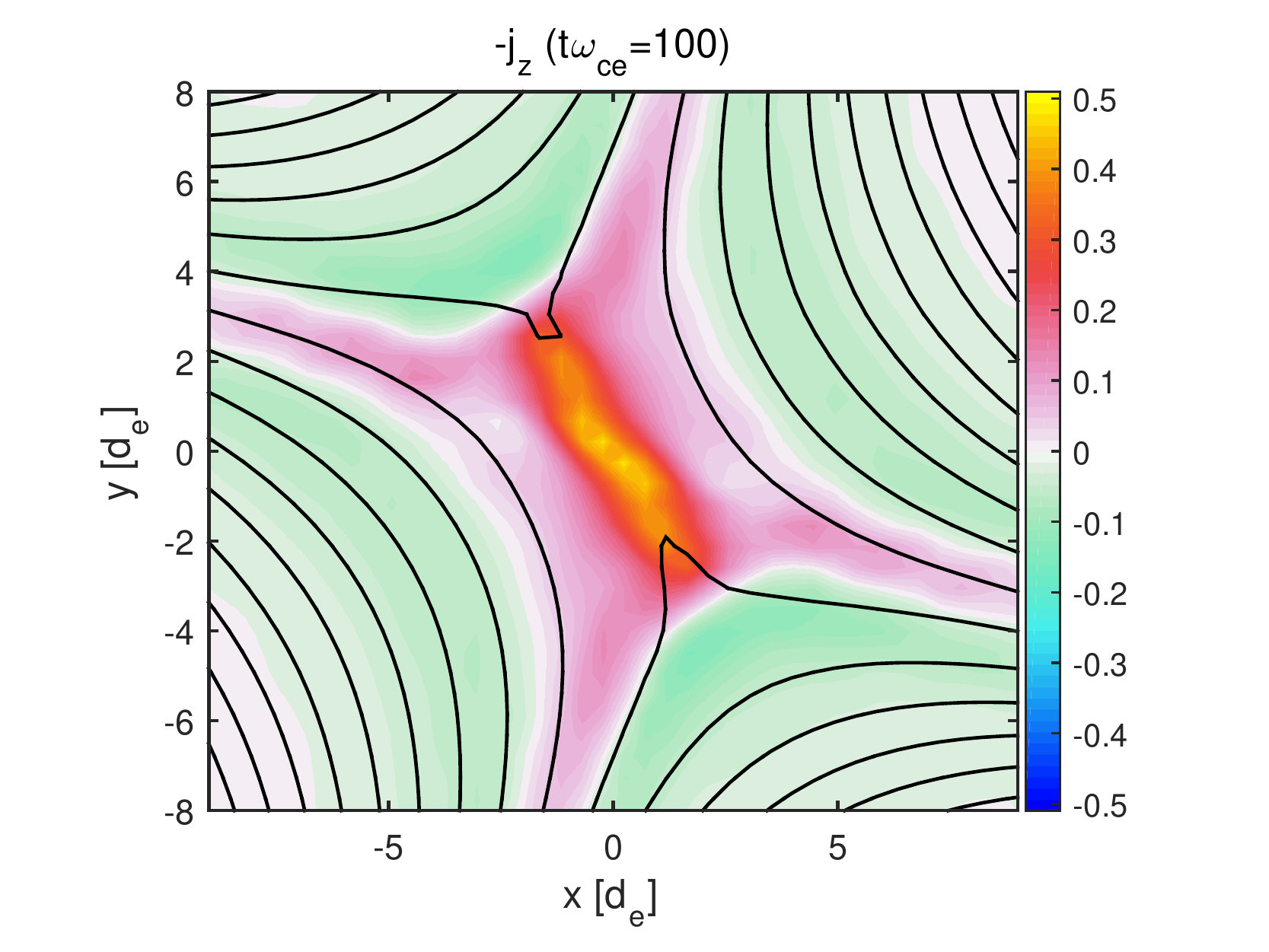}
                \put(-145,128){(a)}
                \includegraphics[clip,width=0.49\textwidth,trim=1cm 0.3cm 1cm 0.2cm]{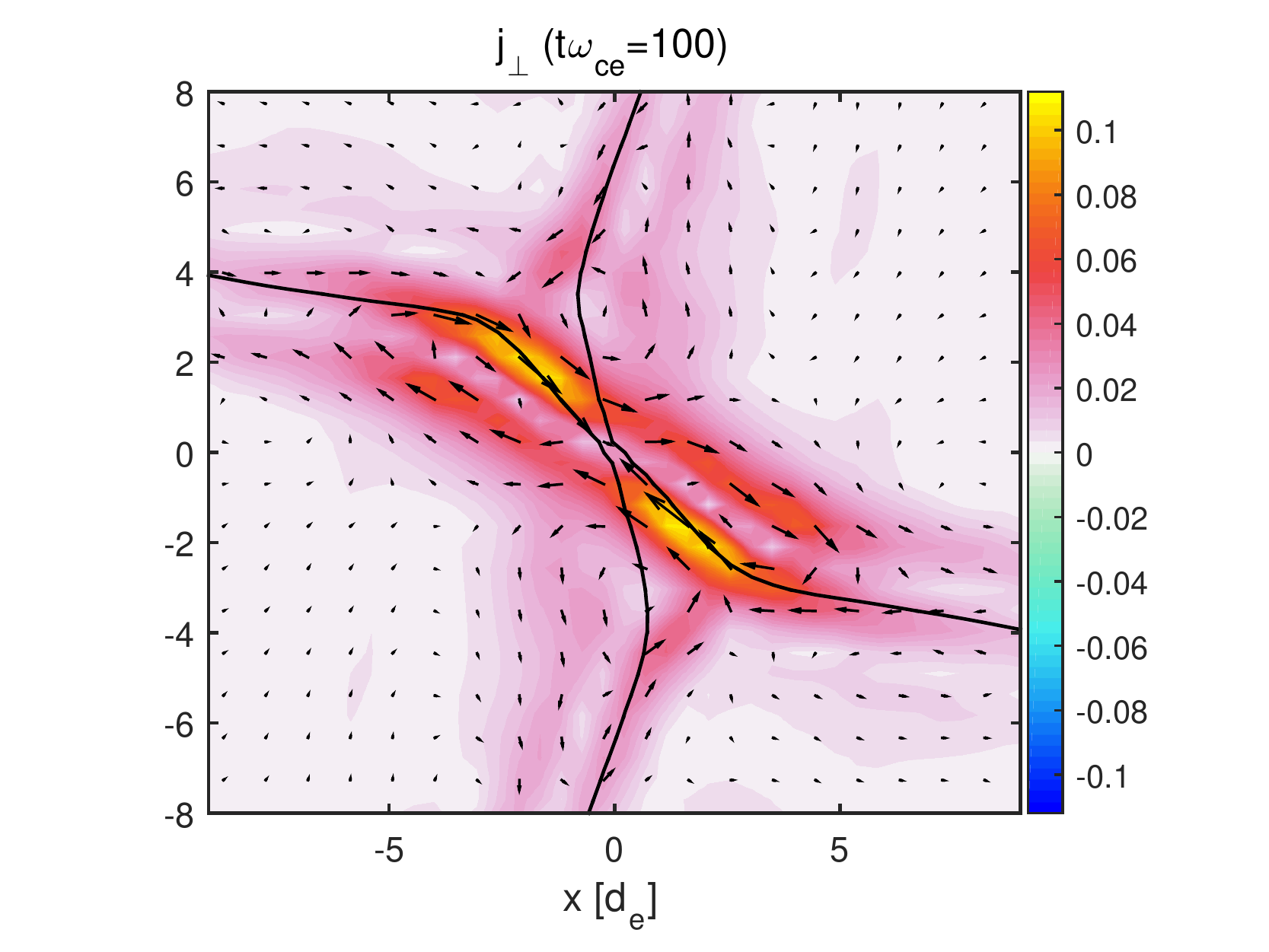}
                \put(-145,128){(b)}
                \caption{
                Color-coded axial (a) and in-plane (b) current density with field lines (black) of the total
                perpendicular magnetic field $\vec{B}_{\perp}=\vec{B}_{\perp}^{p}+\vec{B}_{\perp}^{ext}$,
                where $\vec{B}^p$ is the magnetic field caused by the plasma currents.
                Arrows in (b)  point in the direction of the in-plane current density
                (Reproduced from~\cite{Jain:2017-Vineta} with the permission of AIP Publishing).
                }
                \label{fig:jz_slice}
        \end{figure}
\end{center}

\begin{center}
        \begin{figure}[!ht]
                \includegraphics[clip,width=0.49\textwidth,trim=1cm 0.3cm 0.1cm 0.2cm]{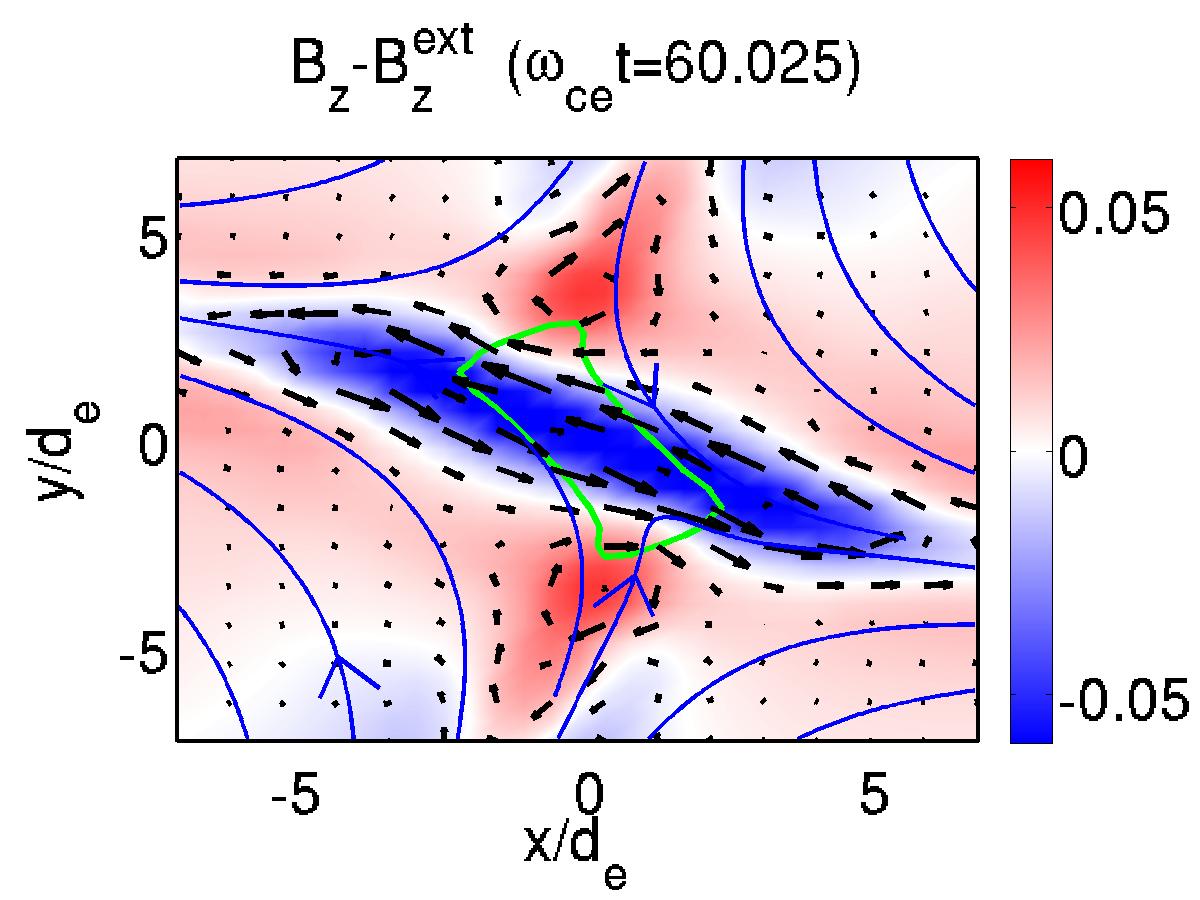}
                \put(-145,110){(a)}
                \includegraphics[clip,width=0.49\textwidth,trim=1cm 0.3cm 0.1cm 0.2cm]{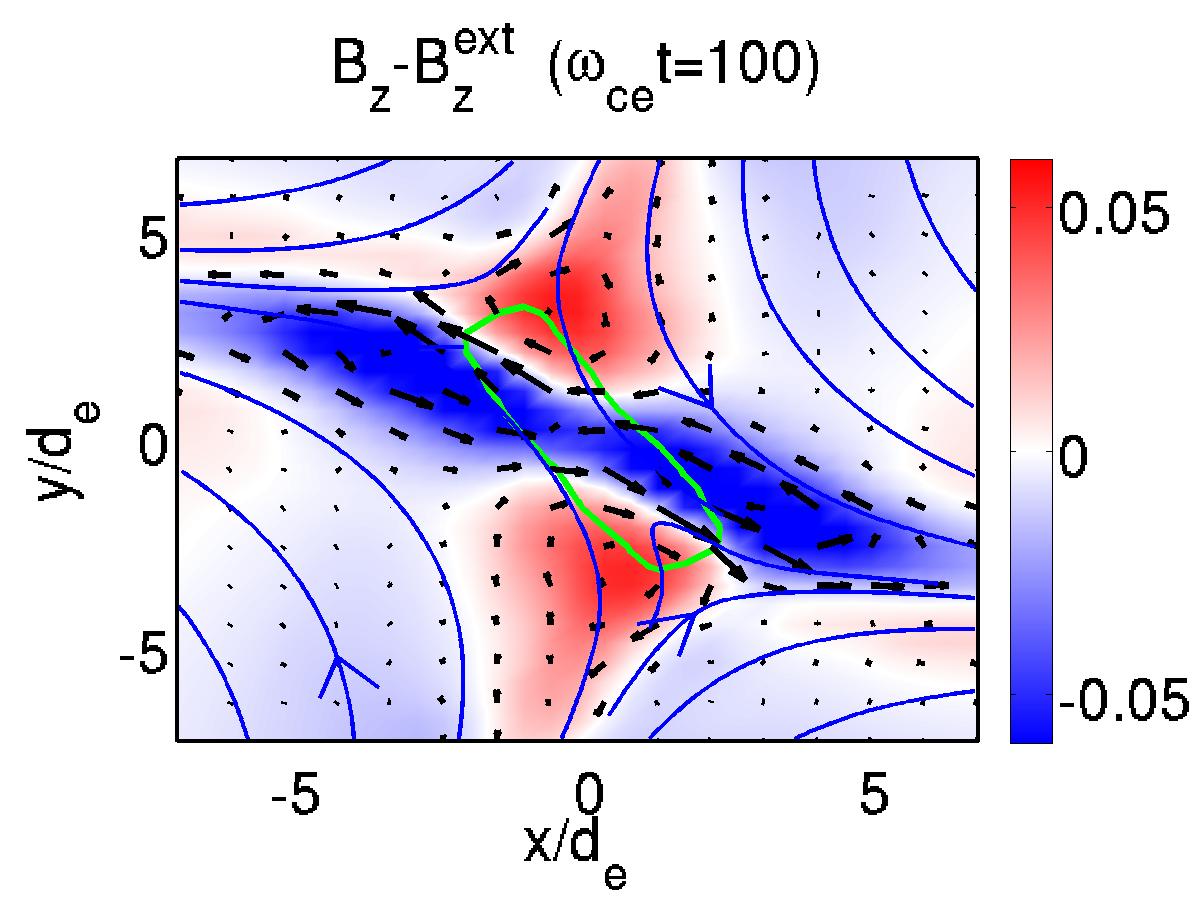}
                \put(-145,110){(b)}
                \caption{Out-of-plane component of the plasma magnetic field, $\hat{z}\cdot(\vec{B}-\vec{B}^{ext})$ (color),
                at two times with projection of magnetic field lines (blue) and electron flow vectors (arrow) in an x-y plane
                The green curve is the contour of the axial current density at $j_z=0.42\, j_z^{max}$.
                     (Reproduced from~\cite{Jain:2017-Vineta} with the permission of AIP Publishing).
                }
                \label{fig:bz_slice}
        \end{figure}
\end{center}

\begin{center}
  \begin{figure}[!ht]
    \includegraphics[width=0.45\textwidth]{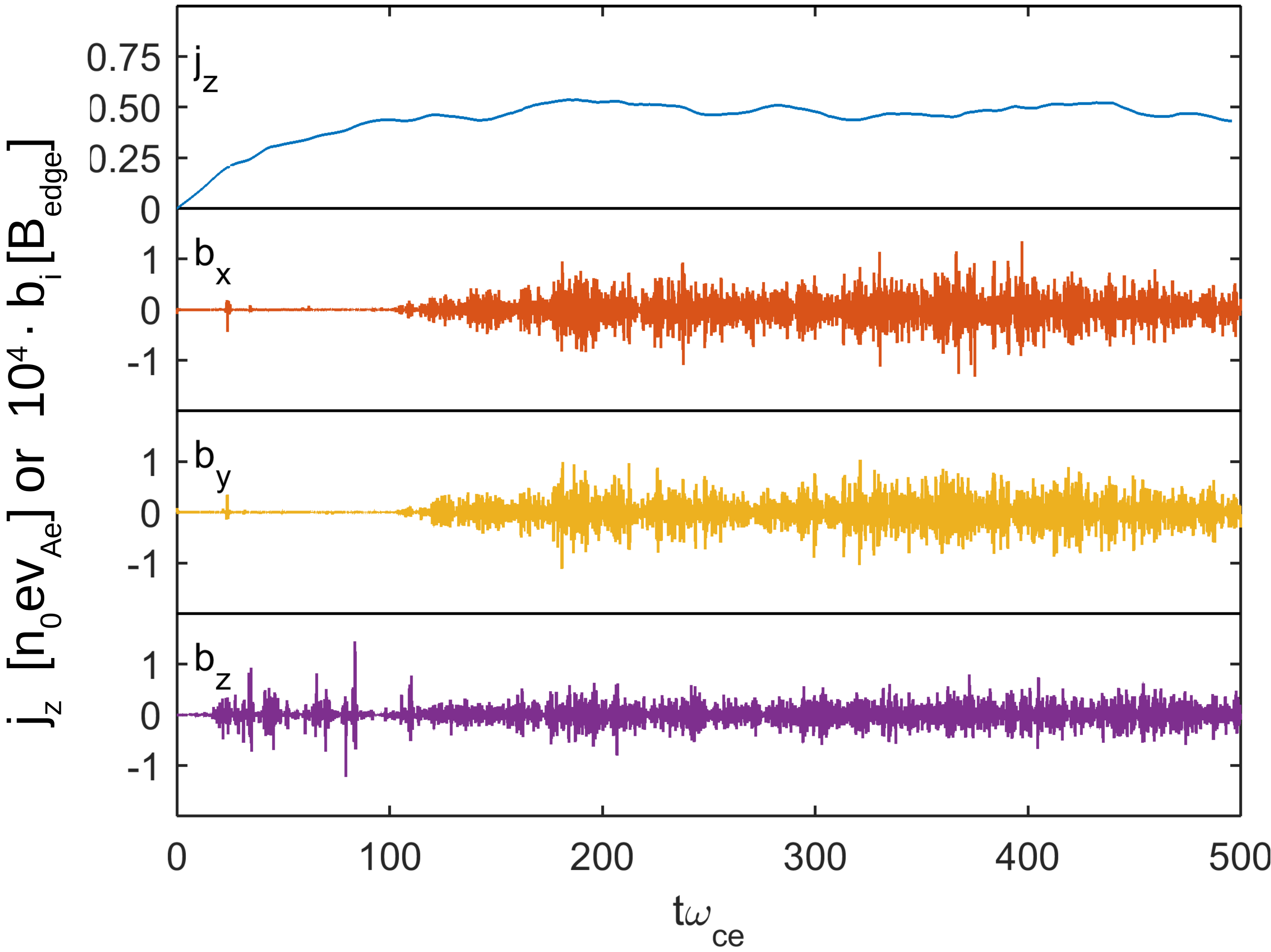}
    \put(-125,115){(a)}
\includegraphics[clip,width=0.45\textwidth]{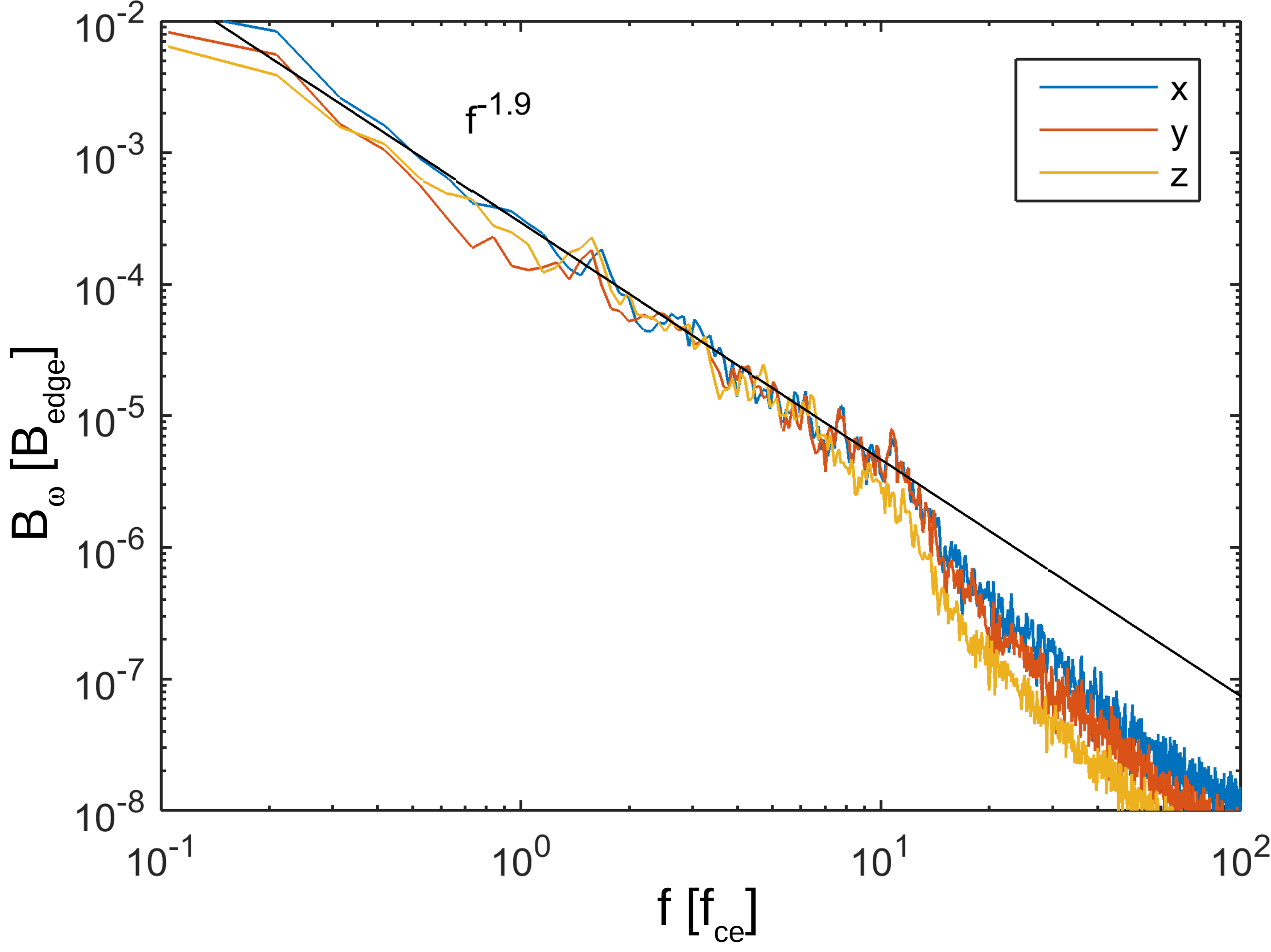}
\put(-125,115){(b)}\\
\includegraphics[clip,width=0.49\textwidth]{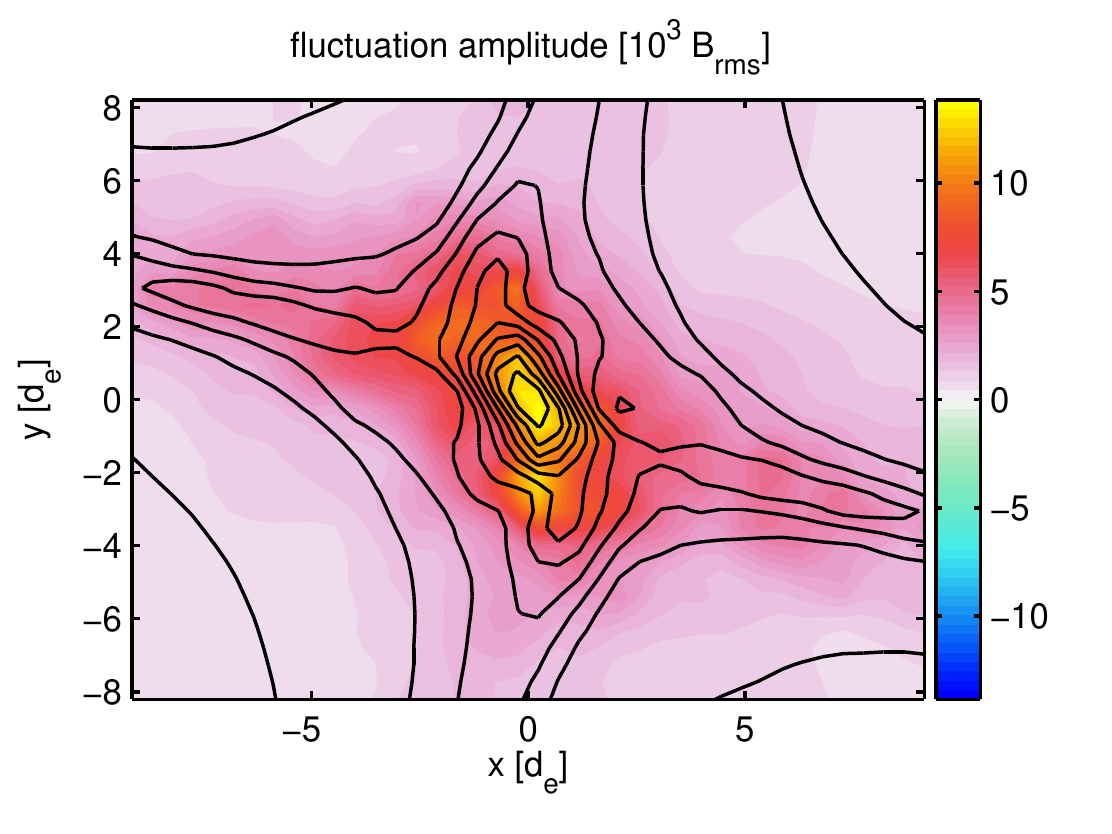}
\put(-140,115){(c)}
\includegraphics[clip,width=0.49\textwidth]{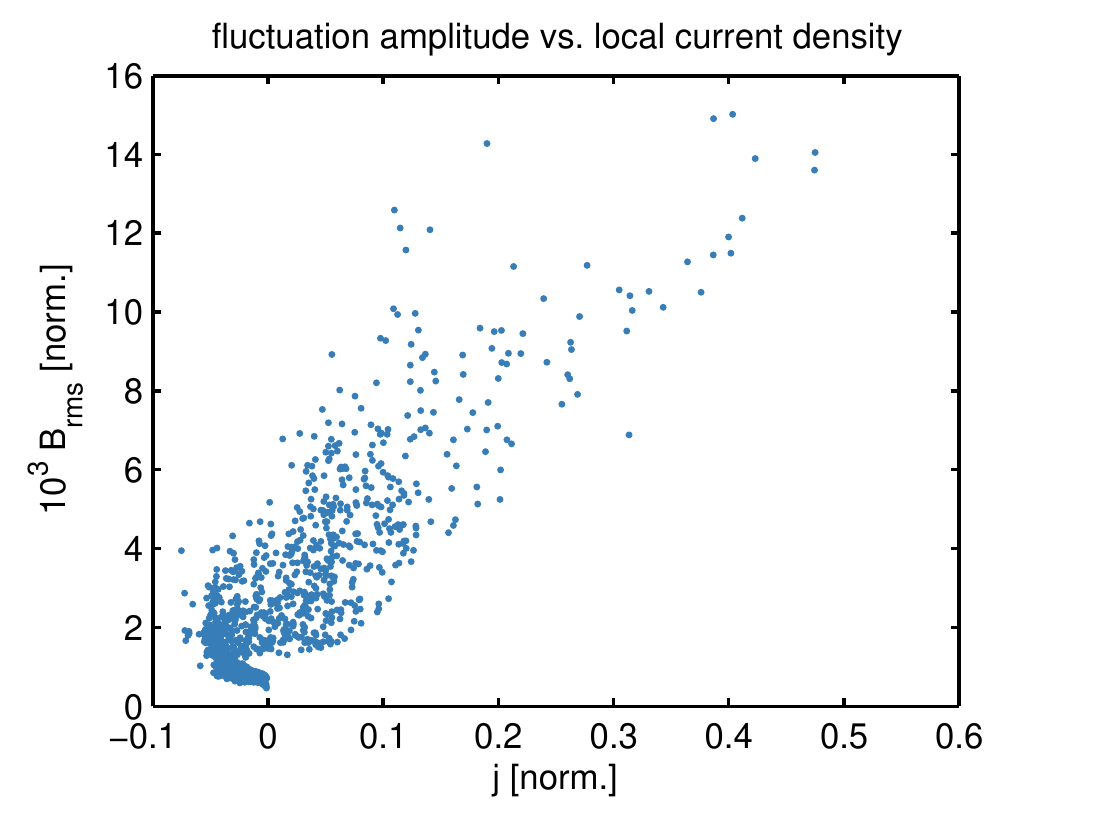}
\put(-150,115){(d)}
\caption{(a) Time evolution of the current density $j_z$ (top panel) and fluctuations in magnetic field components,  $b_i$, ($i=x, \, y, \, z$, bottom three panels), obtained by  high pass filtering ($f_{-3\,\mathrm{dB}}=f_{ce}$)  the plasma magnetic field near the X-line. (b) Amplitude spectra of the $x$, $y$ and $z$ components of the plasma magnetic field $\vec{B}^{p}=\vec{B}-\vec{B}^{ext}$ at the X-point  over two decades of frequencies ($f/f_{ce}=0.1-10$ or $f/f_{ce,g}\approx 0.005-0.5$). (c) Root-mean-square (RMS) fluctuation amplitude (color) and contours of axial current density. (d) Scatter plot of  RMS fluctuations amplitudes with the axial current density over space. The Fourier (in b) and RMS (in c and d) amplitudes were obtained from the magnetic field data in the time interval  $200 <\omega_{ce}t<500$.
(Reproduced from~\cite{Jain:2017-Vineta} with the permission of AIP Publishing).
} \label{fig:spectra}
\end{figure}
\end{center}

The three dimensional EMHD simulations have shown that the evolution of the system remains
two-dimensional ($\partial/\partial z\approx 0$) during the simulation (electron-) time even
for very large simulation boxes in the z-direction.
Figs. \ref{fig:jz_slice} and \ref{fig:bz_slice} show the magnetic fields and currents
in an x-y plane.
In a small disc around the X-point where the in-plane (perpendicular to the guide magnetic field) magnetic field is vanishingly small, electrons are accelerated by the $E_z^{ext}$ field along the $z$-direction. Fig. \ref{fig:spectra}a shows that the current density $j_z$ at the X-point grows until $\omega_{ce}t \approx 100$ and then saturates around $j_z=0.5\, n_0ev_{Ae}\approx 118$ kA/m$^2$, which is of the same order of magnitude as the peak experimental current density (40 kA/m$^2$).
The in-plane drift velocity of electrons, primarily given by $\vec{v}_{e\perp}=\vec{E}_{\perp}\times \hat{z}/B_z$,
modifies the initially ($\omega_{ce}t < 10$) disc-shaped cross-section of $j_z$, respectively stretching and pinching the current channel along the directions parallel and perpendicular to the line connecting the conductors. Simultaneously, $j_z$ extends towards the separatrices. The resulting structure of the out-of-plane current sheet at $\omega_{ce}t=100$ is shown in Fig. \ref{fig:jz_slice}a.

Similar to the experiments, the in-plane electron velocity  in the simulations develops a vortical structure. The vortical structure, however, does not align with the current sheet, unlike the experiments.
The single vortex formed by $\omega_{ce}t=60$  (Fig. \ref{fig:bz_slice}a)
breaks into two vortices at $\omega_{ce}t=100$ (Fig. \ref{fig:bz_slice}b and \ref{fig:jz_slice}b)  due to  the
growth of an electron shear flow instability.  Linear instability analysis revealed that the dominant instability grows on the shear of the in-plane electron flow and requires a finite electron inertia~\cite{jain2017}. In the non-linear state ($\omega_{ce}t > 100$), the instability generates fluctuations in the perpendicular (to the guide field)  magnetic field components. The onset of these fluctuations  coincide with the saturation of the growth of the axial current density, suggesting that the fluctuations in our simulations with no collisional resistivity can provide some kind of anomalous dissipation (see Fig. \ref{fig:spectra}a ).
EMHD simulations of the same setup without electron inertial terms show that the axial current density at the X-point continues to grow
without saturating. Furthermore, the shear flow structures of the out-of-plane and in-plane flows, as shown in Figs. \ref{fig:jz_slice} and \ref{fig:bz_slice}, did not develop. This suggests that electron inertia plays an important role not only in the development of the instability and fluctuations but also in the formation of the shear flow structure on which the instability grows.
Note that the electron inertia allows stable simulations
even in the absence of  resistivity.

The amplitude spectra of the plasma magnetic field fluctuation obtained near the reconnection
X-line, Fig. \ref{fig:spectra}b,  show a clear power law with a spectral index of $\alpha\approx1.9$
for all the components across two decades of frequencies ($f/f_{ce}=0.1-10$  or $f/f_{ce,g}\approx 0.005 - 0.5$, where $f_{ce}=2\pi/\omega_{ce}$ and $f_{ce,g}=2\pi/\omega_{ce,g}$ are the electron cyclotron frequencies in the edge and guide magnetic fields, respectively).
The experimental spectra show a broadband power law behavior over a wider frequency range, extending below the lower hybrid frequency $f_{LH,g}$ in the guide magnetic field. The part of the spectrum at or below $f_{LH,g}$, however,  cannot be studied  in EMHD simulations since the ion motion is not included in the model.
The spectral index $\alpha \approx 1.9$ obtained by EMHD simulations in the
high frequency part ($f > f_{LH,g}$) should be compared to the value $\alpha \approx 2.4$
obtained in the experiment.
As in the experiment, the root-mean-square (RMS) values of the magnetic field fluctuations
calculated for the frequencies below the electron cyclotron frequency correlate well with
the local current density and peak at the center of the current sheet (Fig. \ref{fig:spectra}c).
Fig. \ref{fig:spectra}d shows a good correlation between the two quantities with a nearly linear relationship, as found in the experiment.

Hence, the EMHD model reveals already some consequences of the finite
electron inertia for guide-field magnetic reconnection but fails to describe others,
which are associated with the coupling of the electron fluid to the ion motion
lasting longer than the typical electron-time scales.
The inclusion of ion effects requires hybrid-kinetic models with finite
electron mass which consider also the mobility of the ions.

\subsubsection{Hybrid-kinetic simulation of guide-field reconnection with finite electron mass
and mobile ions
\label{sec:mobile_ions}}

A hybrid-kinetic code with an inertial electron fluid like CHIEF allows to investigate the
influence of the ions in addition to that of the massive electrons.
Let us demonstrate the consequences of mobile ions for guide field reconnection in
an initial setup similar (but not identical) to that of EMHD simulations in section~\ref{sec:immobile}.
Different from the EMHD simulation setup, the spatial position of the current-carrying
wires generating the perpendicular magnetic field are now located
on the y-axis, i.e. rotated by 30$^{\circ}$ with respect to their positions in the
EMHD simulations discussed in section~\ref{sec:immobile}.
The other simulation parameters are:
$r_{\rm gun}=24$ mm, $k_BT_e=6$ eV,  $T_e/T_i=60$, ion plasma beta $\beta_i=0.002$ based on the guide magnetic field $B_g=8.8\,$mT, and electron-ion collision frequency $\nu/\Omega_{ce}=0.01$, with $\Omega_{ce}$ the electron cyclotron frequency based on the guide magnetic field.
The ratio of the guide magnetic field ($B_g$) to the upstream magnetic field ($B_{\rm edge}$)
is $11$.
The simulation box is two-dimensional with a physical size $10 \times 10$ cm and a grid
resolution  $\Delta x/d_e=0.25$. The number of particles is 512 protons per cell.
A realistic ion-to-electron mass ratio is used. The physical electron inertial length corresponds to $d_e=0.26$ cm.

\begin{figure}[!ht]
\includegraphics[width=0.99\linewidth]{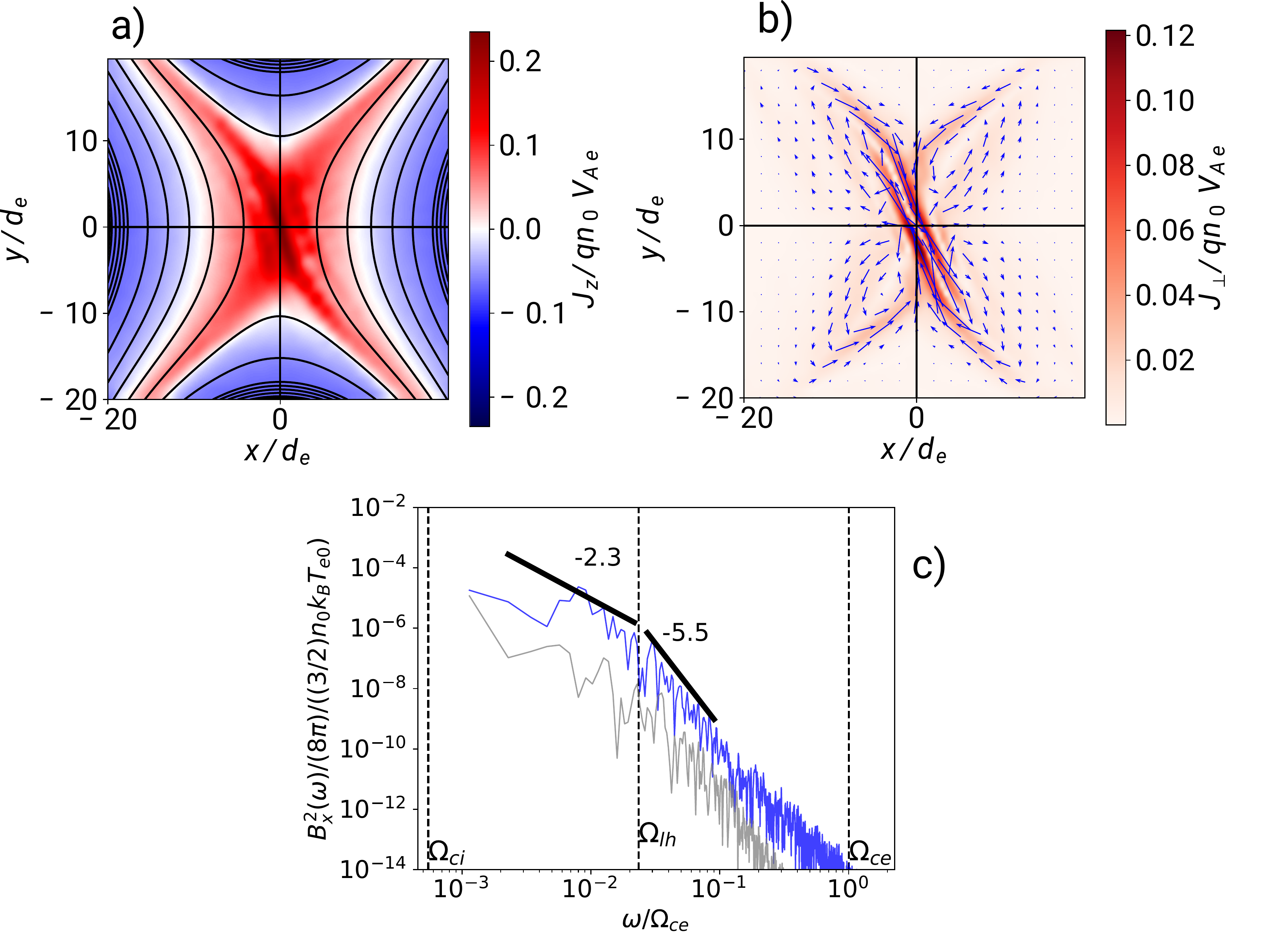}
\caption{Hybrid-kinetic simulation results for the conditions of 
the VINETA.II experiment. Contours plots of a) out-of-plane current
density $\vec{J}_z$  and b) magnitude of the in-plane current density with the arrows representing the vector flow $\vec{J}_{\perp}=\vec{J}_{x}\hat{x} + \vec{J}_{y}\hat{y}$. Both quantities were obtained at $t = 12100\Omega_{ce}^{-1}$ (based on the guide magnetic field). The normalization value $V_{Ae}$ is with respect to the upstream magnetic field. c) Frequency spectrum of in-plane magnetic fluctuations at the X-point (blue lines) and at a point outside of the current sheet (gray line). The oblique straight lines represent an exponential fitting with the indicated spectral slopes below and above the lower-hybrid frequency $\Omega_{lh}$.}
\label{fig:contours1}
\end{figure}

Fig.~\ref{fig:contours1} shows that the profile of the out-of-plane current density (panel a) exhibits a
slight asymmetry along one separatrix arm that can be attributed to the asymmetric ion density due to the presence of the strong guide field. The in-plane current density in Fig.~\ref{fig:contours1}b displays a strong current flow along the separatrix. Unlike the EMHD simulations, it is aligned with the current sheet. This is in agreement with the VINETA.II experiment.
Similar to the experiments, the power spectrum of the in-plane magnetic fluctuations during a given interval in steady-state
($\Delta t \Omega_{ce}=[12100,17600]$)
in Fig.~\ref{fig:contours1}c now shows a spectral break near the lower-hybrid frequency $\Omega_{lh}$ (based on the guide magnetic field). For frequencies higher than $\Omega_{lh}$ the power spectrum steepens indicating a change in the nature of the electromagnetic fluctuations. The power  spectrum has a spectral index $\alpha=5.5$ for frequencies higher than $\Omega_{lh}$ and $\alpha=2.3$ for frequencies lower than $\Omega_{lh}$.
Those values have to be converted to amplitude spectral indices, which results in $\alpha=1.15$ for $\omega<\Omega_{lh}$ and  $\alpha=2.75$ for  $\omega>\Omega_{lh}$.
Note that the corresponding experimental values are $\alpha=1.4-1.8$ and  $\alpha=2.4$, respectively. Although these values still slightly differ from the experimentally obtained ones, they are closer
than those obtained  from EMHD simulations due to the consideration of the  mobile ions.

\subsection{Collisionless plasma turbulence and current sheets
\label{sec:turbulence}}

The turbulent transfer of energy from macroscopic to microscopic kinetic scales such as Larmor radii and inertial lengths of plasma particles,
where it is finally dissipated into heat, is one of the most viable mechanism of energy dissipation 
in collisionless space and astrophysical plasmas.
At kinetic scales, current sheets (CSs) with thicknesses ranging from
ion to electron scales are observed ubiquitously in space observations
of collisionless turbulent plasmas.
A number of observational and simulation studies
strongly suggest that the dissipation of plasma energy is localized in
and around such kinetic scale CSs.

To understand the dissipative processes in current sheets, two-dimensional hybrid-PIC simulations
were performed with massless electrons. These simulations
show the formation of ion scale CSs and
reveal  that they are formed primarily
by electron shear flows with an electron bulk flow velocity exceeding
by far the ion bulk velocity while density variations are relatively small
($<$ 10\%)~\cite{jain2021}.

Current sheets, however, tend to thin down below the ion inertial length.
In hybrid-kinetic plasma models neglecting the electron mass,
CSs thin down all the way to the numerical grid scale where the
thinning is stopped by numerical effects at the grid scale~\cite{azizabadi2021}.
In real collisionless plasmas, of course, the CS thinning would be
stopped by some physical effect at scales below the ion inertial
length, like the electron inertial length and/or the electron
gyro-radius, a finite-electron-mass effect. Then the CSs may become susceptible to electron shear flow
instabilities which can generate the electromagnetic fluctuations observed in space and laboratory experiments  (see section \ref{subsec:vineta}).
The influence of the CS thickness on the turbulence, therefore,
has to be studied by taking into account the electron inertia.

 Hybrid-kinetic simulations with finite electron mass can be used to
 study these processes. Let us illustrate this by the results of
two-dimensional CHIEF-code simulations initialized with randomly
uncorrelated Alfv\'enic fluctuations,

\begin{align}
\vec{B}& = \sum_{k_z=-Nk_z^m/2}^{Nk_z^m/2}\;\sum_{k_y=-Nk_y^m/2}^{Nk_y^m/2}\frac{B_{rms}\sqrt{2}}{k_{\perp}N}(k_z\hat{y}-k_y\hat{z})\cos(k_z z + k_y y + \phi_b) + B_0\hat{x} \\
\vec{V}& = V_A\sum_{k_z=-Nk_z^m/2}^{Nk_z^m/2}\;\sum_{k_y=-Nk_y^m/2}^{Nk_y^m/2}\frac{B_{rms}\sqrt{2}}{B_0k_{\perp}N}(k_z\hat{y}-k_y\hat{z})\cos(k_z z + k_y y + \phi_v)\,\,\, ,
\end{align}

where $\phi_b$ and $\phi_v$ are independent random phases.
For definiteness an amplitude $B_{rms}/B_0=0.24$ and a number
of modes  $N=6$ were chosen.
The minimum wave numbers allowed in the simulation domain along y- and z-directions are $k_y^m=2\pi/L_y=0.061 d_i^{-1}$ and $k_z^m=2\pi/L_z=0.061 d_i^{-1}$, corresponding
to the simulation domain length, $L_y=102.4 d_i$ and $L_z=102.4 d_i$, in the two directions, respectively. The maximum perturbed wave numbers along any of the two directions are
$0.18 d_i^{-1}$.
Note that the total magnetic energy of the magnetic perturbation is equal to the bulk flow
kinetic energy of the velocity perturbations.
A plasma beta based on the out-of-plane background magnetic field of $\beta_i=\beta_e=0.5$ is chosen and an ion-to-electron mass ratio
of $m_i/m_e=25$. The spatial simulations domain is covered by $1024\times 1024$ grid cells
with a grid cell size equal to $\Delta y=\Delta z=0.5 d_e=0.1 d_i$, where $d_e$ ($d_i$) is the electron (ion) inertial length. Five hundred macro-particles (protons) per cell are launched.
The time step is $\Delta t=0.25 \omega_{ce}^{-1}=0.01\omega_{ci}^{-1}$, where $\omega_{ce}$ ($\omega_{ci}$) is the electron (ion) cyclotron frequency,
so that the CFL condition for electron Alfv\'en wave speed is satisfied to $V_{Ae}\Delta t/\mathrm{min}(\Delta y, \Delta z)=0.5$. The plasma resistivity is zero.

Inertia-less-electron hybrid-kinetic simulations are also carried out to compare
with the results obtained for the inertial electrons with the same parameters except the time step which has to be chosen much smaller (by a factor of five,
$\Delta t=0.002\omega_{ci}^{-1}$) since one now has to stabilize
whistler wave instabilities at the grid scale for inertia-less electrons.
Boundary conditions are periodic in all directions.

\begin{figure}[!ht]
\centering
         \begin{minipage}[b]{0.32\textwidth}
                 \includegraphics[width=0.99\textwidth]{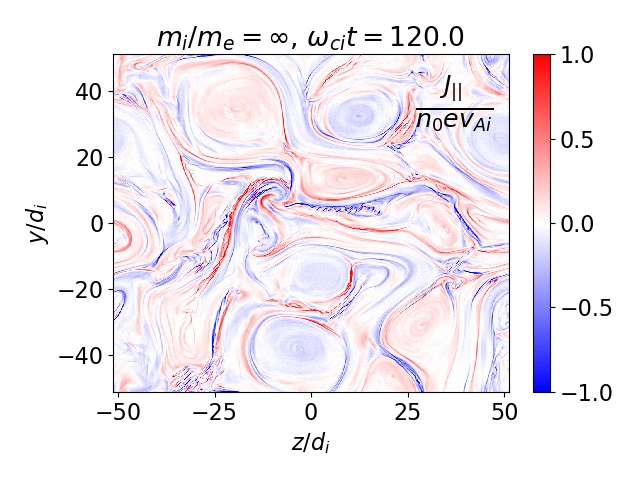}
         \end{minipage}
         \begin{minipage}[b]{0.32\textwidth}
                 \includegraphics[width=0.99\textwidth]{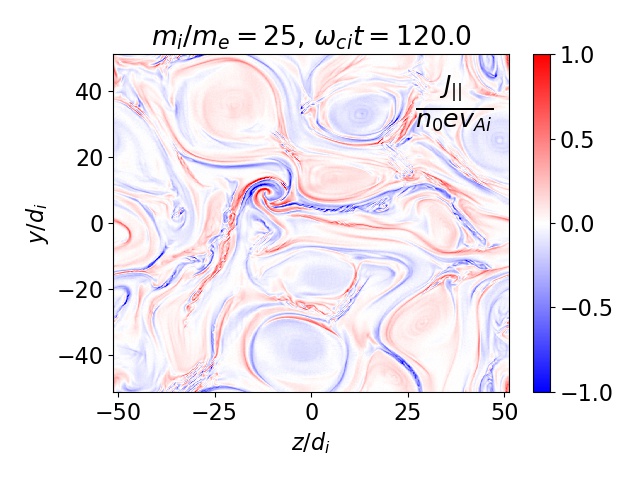}
         \end{minipage}
         \begin{minipage}[b]{0.32\textwidth}
                 \includegraphics[width=0.99\textwidth]{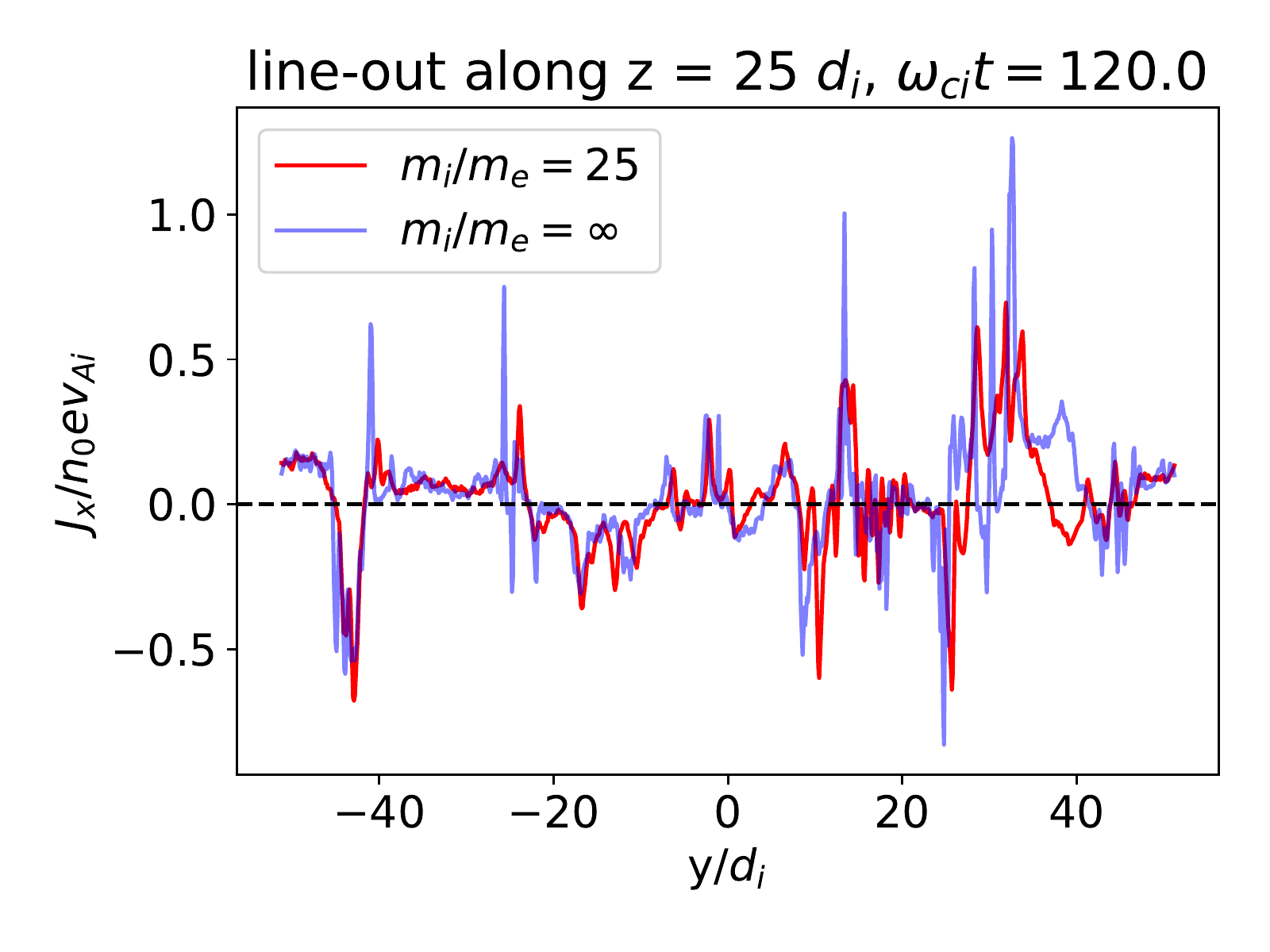}
         \end{minipage}
\caption{Contours of out-of-plane current density for inertia-less (left panel) and inertial (middle panel) electrons 
and comparison of their line-outs along $z=25 d_i$ at $t=120\;\omega_{ci}^{-1}$, the time of maximum turbulent 
activity (right panel). Reprinted from \cite{Jain2022}, with the permission of AIP Publishing. 
\label{fig:standard}}
\end{figure}

Results are shown in Fig.~\ref{fig:standard}. Its two left panels depict
iso-contours of the out-of-plane current density obtained for inertia-less
(left panel) and for the inertial-electron-fluid (middle panel) simulations.
The plots are obtained at $\omega_{ci,B_0}t=120$ --- the time of maximum
turbulence, i.e. when the RMS value of the current density peaks.
While the current densities are quite similar in both cases they are,
however, different at the finer scales.
This can be seen in their line-outs taken along the line $z/d_i=$25
(see Fig. \ref{fig:standard}, right panel). While the current densities in both models are more or less the same
at larger scales they differ at shorter scales.
The current density for the case of the inertia-less electrons is noisier
and spikier than that for obtained with inertial electrons.
The magnitude of the current density spikes are smaller if the
electron inertia is taken into account.
The finite electron inertia interferes with the thinning process
of the CSs at scales $k_{\perp}d_e \sim 1$ before numerical effects (the grid scale) can. This leads to thicker current sheets and smaller current density spikes for inertial electron fluid while the electron bulk flow velocity exceeds the
ion bulk velocity in the CSs for both the inertia-less and inertial electron fluid. 

Electron inertial effects further contribute to the dissipation of the
turbulence via triggering current sheet instabilities. The way the
inertial-electron instabilities affect the dissipation in the CSs embedded
in collisionless plasma turbulence is, however, still an open question to
be investigated.

\subsection{Collisionless shocks}
\label{sec:shocks}

So far collisionless shocks have been simulated mainly by
fully-kinetic PIC-codes or hybrid-kinetic codes with massless
electrons.
The shock front needs to be sustained by dissipation which 
in collisionless plasmas is usually provided by wave-particle
interactions due to the plasma instabilities.
Hybrid codes with massless electrons do not describe the electron scale 
instabilities which may provide the dissipation in particular shock
configurations.

For example,  high-Mach number quasi-perpendicular shocks are known
to be dominated by whistler waves with wave-numbers much larger
than the inverse of the ion skin depth and the steep gradient scales.
Electron inertial effects were taken into account in 
hybrid code simulations of shock waves (see, e.g., chapter 8
of~\cite{Lipatov2002}).
One of the findings of Ref.~\cite{Lipatov2002} for quasi-perpendicular shocks
with Mach numbers greater than six is that the periodic reformation of the
shock front is accompanied by a standing whistler wave. While  the
shock front reformation was already described by hybrid codes with
massless electrons~\cite{Quest1986}, the formation of standing whistler
waves brought in the new physical insights into the breaking of the shock
front after its steepening. The whistler waves also cause additional
dissipation of the energy of the particles crossing the shock front.
Finally, it could be shown this way that the thickness of the shock ramp is
 comparable with the largest whistler wavelength in the system which
 becomes smaller for larger Mach number shocks.
Note that hybrid-kinetic shock simulations with electron
inertia require, therefore, a relatively fine grid in order to resolve
the electron skin depth.
Although this makes simulations more expensive, it also allows to
smooth out short-wavelength electric fields and stabilized
the simulations over longer periods of time compared to
standard hybrid code simulations with massless electrons.
This effect is even more expressed the larger the mass
ratios are.

Another relevant electron inertial effect, also recognized 
by Ref.~\cite{Lipatov2002} (see their chapter 8), is related to the 
ion acceleration in low-beta supercritical quasi-perpendicular
shocks.
In order to reproduce properties of pickup ions as often
observed in planetary magnetosphere, an initial ion ring
distribution is assumed.
Different from the standard diffusive shock acceleration, 
the shock surfing acceleration is active mainly in
quasi-perpendicular shocks.
It is due to the formation of a cross-shock potential
caused by the different masses of electrons and ions.
Ions become trapped between the cross-shock potential
and the upstream magnetic field, drifting in this way along
the convective electric field becoming energized.
This affect mainly the thermal ions, which then become
accelerated to the energies that are needed
to participate in the standard diffusive shock acceleration.
It contributes, therefore, to the solution of the injection
problem of the standard diffusive shock acceleration
models.
A critical ingredient for efficient shock surfing	acceleration
is a very thin shock front.
It was  found that for shock surfing
acceleration of hydrogen and helium ions,
	a shock front on the order of the electron skin depth
	is needed~\cite{Lipatov2002}.
	This finding emphasizes the importance of the electron inertia
	in shock acceleration processes involving different ion species.

Note that only hybrid codes with electron 	inertia, or fully kinetic
PIC codes which are, however, numerically much more expansive,
can reproduce these processes.

\subsection{Global magnetospheric hybrid-code simulations}
\label{sec:global}

After initially global simulations of whole magnetosphere 
were carried out within the framework of MHD models the question
arose, how to include the physics of magnetopause, magnetotail,
reconnection and shock waves as well as the coupling of magnetosphere
and ionosphere.
First hybrid-PIC code simulations were utilized to include ion kinetic
effects of reconnection and shock physics but neglecting the
electron inertia for example by Refs.~\cite{Palmroth2018,Karimabadi2014,Karimabadi2006}
(see also references therein).

The inclusion of the physics due to the finite electron mass in global
simulations is a challenging task which has first been attempted, e.g., by
Swift in 1996~\cite{Swift1996}.
Using curvilinear coordinates the author considered the effects of the
ionospheric plasma by means of a cold ion fluid component.
His simulations included a term related to the electron inertia
into the generalized Ohm's law but neglected it in other places
like in the cold-ion-fluid momentum equation.
The code completely neglected the contribution of electron pressure.
Simulations using this approach revealed that electron inertia
has a marginal effect on the code stability compared 	to the stability
provided by the ad hoc assumed resistivity due to the electron-ion
friction. It should be noted, however that in these simulations the grid
cell size always exceeded the electron skin depth.

Later the same author(s) applied this numerical model to the substorm onset
problem, again, taking into account the ionospheric-magnetospheric
coupling~\cite{Swift2001}. They found that the observed wave activity
behind the dipolarization front could, indeed, be related to the electron
inertia, which can cause wave steepening and, finally,
contribute to the formation of auroral arcs at electron inertial 
length scales. The same electron inertia, on the other hand,
restricts the maximum accessible current density limiting,
this way, the brightness of auroral arcs. It is not clear yet,
however, which other effects of the electron inertia might be critical for
the global magnetospheric dynamics.

 \section{Future possible improvements of hybrid code algorithms}
\label{sec:improvements}

Note that further improvements might make hybrid code algorithms
more efficient in the future.
The calculation of $\vec{E}^{N+1}$ from Eq. (\ref{eq:E_at_NP1}),
e.g., requires $\vec{u}_e^{N+1}$ and $\vec{u}_e^{N+2}$.
The value of  $\vec{u}_e^{N+1}$ is obtained from Eq. (\ref{eq:ue_full})
using $\vec{u}_i^{N+1/2}$ since $\vec{u}_i^{N+1}$ is not available,yet,
at this point of the calculations.
The value of $\vec{u}_e^{N+2}$, on the other hand, is obtained by
advancing the equations for $\vec{W}$, $\vec{B}$, $\vec{u}_e$
from time step $N+1$ to $N+2$ without updating the ion density and
velocity, i.e., by using $\vec{u}_i^{N+1/2}$ and $n_i^{N+1}$.
This approach is justified if ions do not move much in a single time step,
which is true for phenomena on space and time scales of the order of
or shorter than ion characteristic scales.
In fact, correct results for a variety of test problems in this scale range
are reproduced  by the code CHIEF which implements the algorithm
presented in Sec.~\ref{sec:implementation} (see also~\cite{Munoz2018}).
If the scales range from electron to ion scales and beyond as
it is of interest in certain classes of problems, the method of electric
field update needs to be improved.
Indeed, the electric field can also be obtained by solving an elliptic PDE
which can be obtained by taking the curl of Faraday's law and using
the generalized Ohm's law to substitute
for $\partial \vec{J}/\partial t$.  This equation is,

	\begin{eqnarray}
		\nabla\times\nabla\times\vec{E}&+&\frac{n e^2}{m_e\epsilon_0c^2}\left(1+\frac{m_e}{m_i}\right) \vec{E} =-\frac{ne^2}{m_e\epsilon_0c^2}\left(\vec{u}_e+\frac{m_e}{m_i}\vec{u}_i\right)\times\vec{B}\nonumber\\  &+&\frac{e}{m_e\epsilon_0c^2}\left[m_e\nabla.[n\left(\vec{u}_i\vec{u}_i-\vec{u}_e\vec{u}_e\right)]
		+\frac{m_e}{m_i}\nabla.\underbar{P}_i-\nabla p_e\right] \label{eq:elliptic_E}.
	\end{eqnarray}

Here $\underbar{P}_i$ is the ion pressure tensor which can be obtained from the
second moment of the ion distribution function.
The calculation of $\vec{E}^{N+1}$ from Eq. \ref{eq:elliptic_E} needs $n^{N+1}$, $\vec{u}_e^{N+1}$, $\vec{u}_i^{N+1}$, $\vec{B}^{N+1}$ and $\underbar{P}_i^{N+1}$ but not $\vec{u}_e^{N+2}$. This avoids the need of advancing the equations for $\vec{W}$, $\vec{B}$ and $\vec{u}_e$ by an additional time step without updating ion quantities.
The calculation of $\vec{E}^{N+1}$, however, stays still approximate
as  $\vec{u}_i^{N+1}$ and $\underbar{P}_i^{N+1}$ are not known yet.
But now the problem is reduced to the standard issue of hybrid codes,
i.e. to the refinement of $\vec{E}^{N+1}$ given $\vec{u}_i^{N+1/2}$
for which various methods discussed in chapter 3 of this textbook can be
applied, taking into account the additional deposition of the ion pressure
tensor onto the grid.

\begin{acknowledgement}
The work of  N.J. was financially supported by the German Science Foundation DFG, projects JA 2680-2-1 and BU 777-16-1, the work of P.M. was supported by the DFG projects MU-4255/1-1 and BU 777-17-1.
We further acknowledge the use of the computing resources of the Max Planck Computing and Data Facility (MPCDF) at Garching, Germany and the valuable cooperation with Drs. Markus Rampp and Meisam Tabriz supporting the efficient parallelization the CHIEF code.
We also thankfully acknowledge the possibility of using computing the resources of the Max-Planck Institute for Solar System Research at G\"ottingen, Germany, and at the Technical University Berlin.
\end{acknowledgement}

\end{document}